\documentclass[a4paper,fleqn]{cas-dc}

\usepackage{graphicx}
\usepackage[numbers]{natbib}
\usepackage{pifont}

\newcommand{\cmark}{\ding{51}} % checkmark
\newcommand{\xmark}{\ding{55}} % x
\def\tsc#1{\csdef{#1}{\textsc{\lowercase{#1}}\xspace}}
\tsc{WGM}
\tsc{QE}
\begin{document}
\let\WriteBookmarks\relax
\def\floatpagepagefraction{1}
\def\textpagefraction{.001}

\shorttitle{Quantifying Uncertainty in Battery Storage Lifetime Predictions}    

\shortauthors{M. Graner et al.}  

\title [mode = title]{From Laboratory Aging Studies to Field Predictions: Quantifying Uncertainty in Battery Storage Lifetime Predictions}




\author[1,3]{Melina Graner}[orcid=0009-0007-3333-6853]%[<options>]
\cormark[1]

%\fnmark[1,3]

\ead{melina.graner@tum.de}

\credit{Conceptualization, Methodology, Software, Validation, Formal analysis, Investigation, Data curation, Writing–-original draft, Visualization}

\affiliation[1]{organization={Technical University of Munich, TUM School of Engineering and Design, Department of Energy and Process Engineering, Chair of Electrical Energy Storage Technology (EES)},
            %addressline={}, 
            city={Munich},
%          citysep={}, % Uncomment if no comma needed between city and postcode
            %postcode={}, 
            %state={},
            country={Germany}}

\author[2]{Jan Figgener}%[]

% Footnote of the second author

\credit{Resources, Writing–-review \& editing }

\affiliation[2]{organization={RWTH Aachen University, Institute for Power Electronics and Electrical Drives (ISEA)},
            %addressline={}, 
            city={Aachen},
%          citysep={}, % Uncomment if no comma needed between city and postcode
            %postcode={}, 
            %state={},
            country={Germany}}

\author[1]{Andreas Jossen}%[]
\credit{Supervision, Writing–-review \& editing}

\author[3]{Holger Hesse}%[]
\credit{Funding acquisition, Supervision, Writing–-review \& editing}

\affiliation[3]{organization={Kempten University of Applied Sciences; Institute for Energy and Propulsion Technologies},
            %addressline={}, 
            city={Kempten},
%          citysep={}, % Uncomment if no comma needed between city and postcode
            %postcode={}, 
            %state={},
            country={Germany}}

\cortext[1]{Corresponding author}



\begin{abstract}
Predicting how long a battery energy storage system will last is critical for warranty design, maintenance planning and investment decisions, yet degradation models are mostly deterministic and rarely validated against real field data. We apply an open-source probabilistic degradation framework, combined with a cell-to-system approximation, to bridge the gap between cell-level laboratory aging models and system-level field predictions for residential battery energy storage systems with quantified uncertainty. The framework predicts cell-level state-of-health to within 0.4\% mean absolute error, roughly half the error of prior models for this dataset. When applied to field operation data, the framework's predictions are consistent with all three available system-level capacity measurements —- a benchmark rarely available for open probabilistic degradation models. Cell-level heterogeneity is approximated by two bounding stress scenarios differing only slightly (a 5\% spread in temperature and a 9\% spread in current). The mean degradation trajectories of the two scenarios reach end-of-life 10 months apart, while the full predicted end-of-life range across both scenarios spans approximately three years, about a third of the expected system lifetime. We link this uncertainty to two drivers. One is a systematic mismatch between laboratory test conditions and field-representative operating stress. The other is variability in the training data itself. These insights translate into concrete, resource-efficient recommendations for future aging study design, supporting more confident predictions of battery lifetime under real-world conditions.
\end{abstract}

\begin{graphicalabstract}
\includegraphics[width=\textwidth]{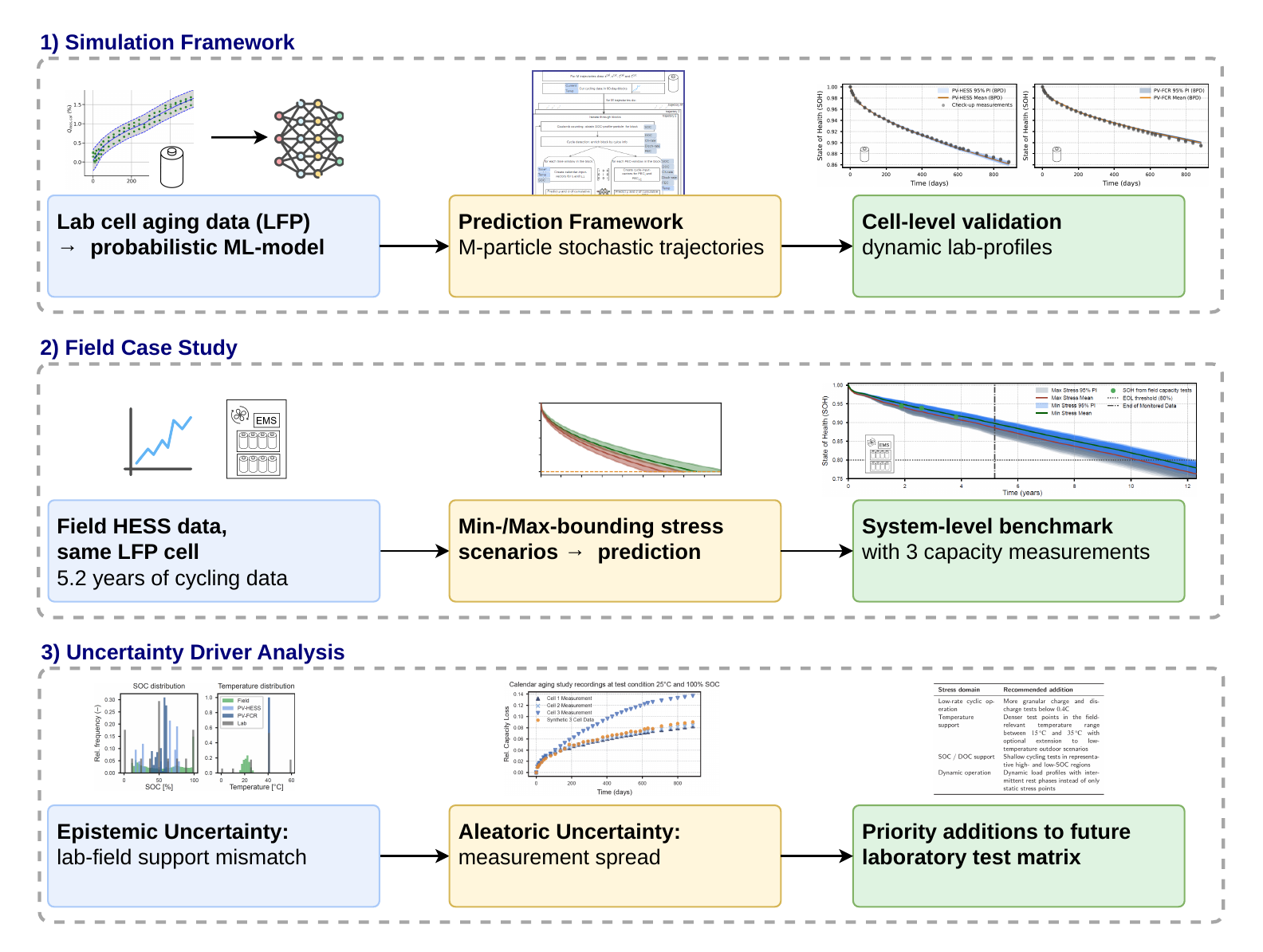}
\end{graphicalabstract}

\begin{highlights}
\item Open-source framework for BESS degradation prediction with quantified uncertainty
\item Cell-to-system approximation successfully benchmarked against BESS field measurements
\item Predicted end-of-life uncertainty amounts to a third of system lifetime
\item Uncertainty driver analysis exposes laboratory-field coverage gaps in test design
\item Recommendations for future resource-efficient, field-relevant aging test campaigns
\end{highlights}

%\nocite{*}

% Keywords


\maketitle

\section{Introduction}\label{intro}
The rising adoption of sustainable energy technologies has driven a sharp increase in demand for efficient and reliable solutions for energy storage. As of today, lithium-ion batteries are the energy storage technology of choice in electric vehicles (EVs), stationary battery energy storage systems (BESS) and consumer electronics, owing to their favorable combination of energy and power density~\cite{GlobalEVOutlook2025.}. Their long-term performance, however, is limited by degradation processes that gradually reduce capacity and power capability~\cite{Birkl.2017}. Battery aging arises from both cyclic aging during charge-discharge operation and calendar aging during storage, each driven by irreversible chemical and structural changes within the cell~\cite{Vetter.2005}. In this rapidly growing market segment with a multitude of new cell types, reliable prediction of degradation trajectories is essential both for BESS manufacturers, who provide warranties guaranteeing a minimum state-of-health (SOH) within a defined period, and for operators, who rely on these predictions for investment and operational planning. This dual relevance has made SOH estimation a well-researched domain.
 
%The rate of battery aging depends on multiple external stress factors such as temperature, current, and depth-of-discharge, allowing operators to influence aging through operational management. 

\begin{table}[t]
%\caption{Table of Symbols}
\label{tab:symbols}
\centering
\setlength{\fboxsep}{6pt}
\footnotesize
\fbox{
\begin{tabular}{@{}p{0.22\columnwidth}p{0.68\columnwidth}@{}}
\multicolumn{2}{l}{\textbf{Abbreviations}} \\[2pt]
BESS  & Battery energy storage system \\
BMS   & Battery management system \\
BOL   & Begin-of-life \\
BPD   & BattProDeep (Git repository~\cite{BattProDeep.github}) \\
DOC   & Depth-of-cycle \\
EMS   & Energy management system \\
EOL   & End-of-life \\
EV    & Electric vehicle \\
FCR & Frequency containment reserve \\
FEC   & Full equivalent cycles \\
HESS  & Home energy storage system \\
LFP   & Lithium Iron Phosphate \\
LMO   & Lithium Manganese Oxide \\
MAE   & Mean absolute error \\
ML    & Machine learning \\
NMC   & Lithium Nickel Manganese Cobalt Oxide \\
PI    & Prediction interval \\
PV    & Photovoltaic \\
RUL   & Remaining-useful-life \\
SOC   & State-of-charge \\
SOH   & State-of-health \\[4pt]
\multicolumn{2}{l}{\textbf{Parameters and Symbols}} \\[2pt]
$L^\mathrm{cal}, L^\mathrm{cyc}$         & Cumulative relative cal./cyc. loss [.] \\
$L^\mathrm{total}$                & Total cumulative loss [.] \\
$\mathbf{x}^{cal}, \mathbf{x}^{cyc}$ & Cal./cyc. feature vector \\
$f_\mu, f_\sigma$          & Mean/std-dev prediction function (cal.)  \\
$g_\mu, g_\sigma$          & Mean/std-dev prediction function (cyc.)  \\
$f^{-1}$                   & Inverse virtual-time prediction function \\
$t^*$                      & Virtual time [h] \\
$T$                        & Temperature [\textdegree C] \\
$w$                        & Cal. time window [h] \\
$w_{fec}$                  & Cyc. FEC window [.] \\
$\Delta t$                 & Time window duration [h] \\
$\Delta FEC$               & Number of FECs in FEC window [.]  \\
$K$                        & Number of bootstrap models [.] \\
$M$                        & Number of  trajectories [.] \\
$\varepsilon^m$            & Aleatoric uncert. contribution parameter [.] \\
$\zeta^m$                  & Epistemic uncert. contribution parameter [.] \\
$\sigma^{L^{\mathrm{al}}}$            & Aleatoric uncert. (std-dev) [.] \\
$\sigma^{L^{\mathrm{ep}}}$           & Epistemic uncert. (std-dev) [.] \\
$I^{min/max,cell/mod}$     & Min./max. current [A] \\
$n^{parallel}$             & Parallel cells per module [.] \\
$T^{max/min,cell}$         & Max./min. cell temp. [\textdegree C] \\
$C_{\mathrm{field}}$       & Measured extractable field capacity [Ah] \\
$C_{\mathrm{BOL}}$         & BOL system capacity [Ah] \\
Ch-rate                     & Charge rate [C] \\
Disch-rate                  & Discharge rate [C] \\
$\Delta SOC$               & Usable SOC range [.] \\
\end{tabular}
}
\end{table}

\begin{figure}[tb]
    \centering
    \includegraphics[width=1\columnwidth]{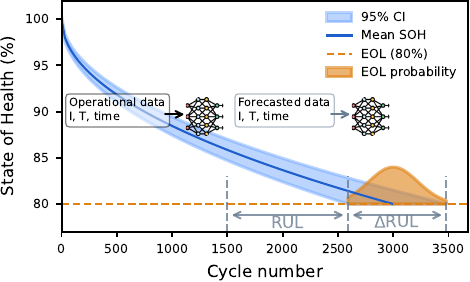} 
    \caption{The abstract formulation of the problem of probabilistic SOH trajectory estimation using probabilistic regression. Future usage data forecasts enable the prediction of future SOH trajectories. If an EOL criterion is defined (in this case 80\% SOH), the RUL can be determined and the likelihood when to expect the EOL can be estimated (EOL probability). }
    \label{fig:problem}
\end{figure}

As illustrated in Fig.~\ref{fig:problem}, operational time-series data such as current, voltage, and temperature can be used to estimate battery SOH. If future usage can also be anticipated, these data further enable the prediction of future SOH trajectories. In combination with a predefined end-of-life (EOL) criterion, such trajectories provide the basis for estimating the remaining-useful-life (RUL) of the battery system. However, most existing degradation models provide only deterministic point predictions and therefore do not reflect the uncertainty associated with long-term aging predictions~\cite{Thelen.2024}. For real-world applications such as maintenance planning, warranty assessment, and lifecycle management, this uncertainty information is essential.

A further challenge is the transfer from cell-level degradation modeling to full battery systems operated in the field. Most existing approaches focus on SOH estimation and degradation modeling at the level of individual cells. By contrast, the transfer of such models to full battery systems remains insufficiently addressed, and system-level datasets with validated field capacity measurements are rare~\cite{Figgener.2024}. This limits the practical applicability of cell-level aging models, since warranty compliance and operational value are ultimately determined at system level rather than at the level of isolated cells.

To address this gap, the present study builds on a previously developed open-source probabilistic framework for battery aging prediction and extends its application toward a more rigorous field-oriented assessment of residential HESS. Deep learning models trained on comprehensive Lithium Iron Phosphate (LFP) cell aging datasets are used to generate uncertainty-aware predictions of capacity-based SOH and RUL.
%We show that those models can be used for probabilistic system-level aging prediction. This enables operators to make early and informed decisions regarding predictive maintenance, warranty management, retrofitting, repurposing, or second-life applications, thereby maximizing both the economic value and environmental benefits of already deployed batteries. Although the emphasis here is on stationary storage systems, the same principles are relevant to all battery-powered devices, including EVs and consumer electronics, where reliable lifetime estimation and safety are critical. 
The key contributions of this paper to the field of battery prognostics can be summarized as follows: 
%\textcolor{blue}{mention the principle of: training predictor for LFP cell dataset and use validation datasets and field datasets for model framework development }
\begin{itemize}
%\item \textbf{Probabilistic Regression Framework:} Formulation of battery aging as a probabilistic regression problem. The framework outputs multiple degradation trajectories, each representing a plausible degradation path. 
%\item \textbf{High-Precision Validation:} Achievement of an absolute 0.4\% SOH prediction error on dynamic validation profiles on cell-level while providing narrow 95\% prediction intervals (PIs). 
\item \textbf{Transfer of cell-level aging models to system-level predictions:} A probabilistic LFP cell aging model is transferred to system-level SOH prediction for residential BESS by means of a topology and stress-scenario-based cell-to-system approximation.
\item \textbf{Validation and benchmarking:} The resulting predictions are evaluated on dynamic cell-level validation profiles and benchmarked against field capacity measurements of residential HESS. The results indicate that the proposed approximation provides a plausible envelope for the observed system-level degradation.
\item \textbf{Uncertainty Driver Analysis:} A quantitative comparison of laboratory and field stress-factor distributions identifies coverage misalignments as a potential explanation for increased epistemic uncertainty. The degree of variability within the model's training datasets is identified as a driver of aleatoric uncertainty.
\item  \textbf{Implications for laboratory aging study design:} Recommendations for improving future laboratory aging campaigns are derived with respect to the tradeoff between accelerated degradation testing and representative coverage of real operating conditions.
\end{itemize}

Beyond their methodological contribution, these contributions carry direct practical relevance: the prediction intervals give BESS manufacturers and operators a quantified basis for warranty design, maintenance scheduling, and second-life planning, while the uncertainty-driver analysis and resulting test-matrix recommendations give laboratories a concrete, resource-aware basis for designing future aging campaigns.

\subsection{State of the Art}
Reliable and transparent SOH estimation for battery systems is becoming increasingly relevant for regulation, warranty assessment, and operational risk evaluation~\cite{Office.}. At the same time, there is still no universally standardized procedure for determining battery states such as state-of-charge (SOC) and SOH in field operation.

The aging behavior of single battery cells has been studied extensively in various laboratory aging campaigns, summarized e.g. in~\cite{Schreiber.2026} and~\cite{dosReis.2021b}. In such studies, cells are exposed to controlled cycling or storage conditions, and degradation progress is monitored through periodic check-ups, typically by measuring capacity and internal resistance. These datasets form the basis for a broad range of degradation models. Early approaches were dominated by empirical and semi-empirical formulations that relate capacity loss or resistance increase to operating conditions and stress factors. In parallel, physics-based and electrochemical models were developed, which describe internal degradation mechanisms in greater detail. While these approaches offer physical interpretability and insight into degradation modes, they often require substantial parametrization effort and are computationally demanding. This can limit their applicability under complex and highly variable operating conditions as well as their transferability across cell types~\cite{Rosewater.2019,LozanoRuiz.2025,Hu.2020,Chen.2024}. 

Against this background, data-driven methods, and in particular machine learning (ML), have gained increasing importance for battery state estimation~\cite{Zhang.2019b}. Advances in learning algorithms, the availability of open-source ML libraries and large-scale battery aging datasets have enabled the rapid expansion of data-driven methods for modeling complex degradation behaviors~\cite{Zhang.2019b}. Their appeal lies in the ability to learn nonlinear relationships directly from degradation data without explicitly modeling the internal electrochemical processes. Depending on the approach, ML models operate either on raw signals such as current and voltage (purely data-driven ML models) or on extracted aging-related features (feature-based data-driven ML models)~\cite{Sulzer.2021}. Contemporary ML applications predominantly address SOH estimation and prediction (based on capacity or internal resistance) and RUL prediction~\cite{Hu.2021,Hasib.2021,Wang.2021}. Beyond these established tasks, T.~Li et al. ~\cite{Li.2024} and W.~Li et al.~\cite{Li.2021c} addressed lifetime prediction and capacity trajectory prediction from limited early-life data. However, despite this methodological diversity, most ML-based approaches still provide deterministic point predictions and do not explicitly represent predictive uncertainty, although such uncertainty is crucial for risk-aware decision-making in practical battery applications~\cite{Thelen.2024}.

A further limitation concerns the representativeness of the available battery aging datasets. The review by Schreiber et al.~\cite{Schreiber.2026} shows that most published laboratory aging studies focus on Lithium Nickel Manganese Cobalt Oxide (NMC) chemistries, whereas only 24\% address LFP cells. This is misaligned with current industry deployments, where LFP chemistry accounts for more than half of EV batteries and more than 90\% of BESS worldwide in 2025~\cite{GlobalEVOutlook2025.}. 

\begin{table*}[ht]
\centering
\caption{Overview of selected publications presenting SOH estimation approaches for real world battery systems, compared with this contribution.}
\label{tab:lit_review_calendar}
\resizebox{1\textwidth}{!}{
\begin{tabular}{llllllll}
\hline
Study    & Chemistry &  Prob.& Metric&  Use case &Training data & Validated with & Publicly  \\
&&&&&origin& field measurm. &available\\
\hline
Song \cite{Song.2020}  & NMC & \xmark & Capacity & EV& Field systems & \xmark & \xmark\\
Aitio \cite{Aitio.2021}  & lead-acid & \cmark & Int. Resist.& BESS & Field systems & \xmark & \cmark\\
Figgener \cite{Figgener.2024} & LMO, NMC, LFP & \xmark & Capacity & BESS& Field systems & \cmark & \cmark \\
Yang \cite{Yang.2024}& NMC & \xmark & Capacity & EV  & Lab. cells & \xmark& \xmark\\
Q. Zhang \cite{Zhang.2021} & LFP & \xmark & Capacity & BESS & Field cells &  \xmark & \xmark \\
J. Zhang \cite{Zhang.2024} & agnostic & \xmark & Av. Energy  & EV & Field and lab. & \xmark &\xmark\\
&&&&&cells and systems&&\\
X. Zhang \cite{Zhang.2018}& NMC & \cmark & Capacity & EV &Lab. systems& \xmark & \xmark \\
Hou \cite{Hou.2026}& agnostic & \xmark & Capacity & EV &Field systems&\xmark & \xmark \\
Hong \cite{Hong.2024}& LFP & \xmark & Int. Resist. & EV &Field systems&\xmark & \xmark \\
She \cite{She.2020}& LFP & \xmark & Capacity & EV &Field systems&\xmark & \xmark \\
Qu \cite{Qu.2026}& LFP & \xmark & Capacity & BESS &Field systems&\xmark & \xmark \\
Pozzato \cite{Pozzato.2023}& NMC & \xmark & Capacity & EV &Field systems&\xmark & \xmark \\
Huo\cite{Huo.2021}& NMC & \cmark & Capacity & EV &Field systems&\cmark & \xmark \\
Wang\cite{Wang.2023b}& LFP & \cmark & Capacity & EV &Field systems&\xmark & \xmark \\
He\cite{He.2021}& LFP & \xmark & Capacity & EV &Field systems&\xmark & \xmark \\
\textbf{This contribution}  & \textbf{LFP }& \textbf{\cmark} &\textbf{Capacity} & \textbf{BESS} & \textbf{Lab. cells} & \textbf{\cmark} & \textbf{\cmark} \\ 
\hline 
\end{tabular}
}
\end{table*}

Single-cell SOH estimation is well-established under laboratory conditions, unlike that of (multi-cell) systems operating in the field. The single cell SOH estimation and prediction models are rarely validated on (system-level) field data and instead, field-operation stress is typically mimicked in the laboratory~\cite{Naumann.2020, Elliott.2020}. Often the approaches are tightly parametrized on cell-level charge and discharge behavior, a parametrization that is not directly reproducible at system level. 
%A variety of public datasets exist that describe the operational behavior of battery systems deployed in EVs  \cite{Pozzato.2023, Rucker.2024, Schreiber.2026} and BESS \cite{Aitio.2021, Figgener.2024} and are used to develop battery state estimators on system-level. 
Table \ref{tab:lit_review_calendar} shows a selection of SOH estimators for real world battery systems such as EVs and BESS across different cell chemistry types (among them LFP, NMC, lead-acid, Lithium Manganese Oxide (LMO)). Most existing approaches focus on EV applications, while comparatively few studies address stationary BESS  deployments, despite their rapidly growing market relevance. The majority of studies use deterministic ML-based approaches, whereas probabilistic SOH estimation remains comparatively underexplored, despite its relevance for warranty assessment, operational risk evaluation, and regulatory transparency requirements. In addition, the training data does not always originate from field system operation, even though system SOH is the target of estimation. Validation against true field capacity measurements is particularly rare, mainly because reference capacity tests require prolonged system downtime and induce operational revenue losses~\cite{Deline.2019,Dubarry.2021}. Consequently, several studies rely on indirect approximations such as partial-cycle ampere-hour integration~\cite{Song.2020, Hou.2026} or check-ups on packs cycled in the laboratory~\cite{Zhang.2024}. Some approaches use internal resistance instead of capacity as a health metric, as it is easier to determine during field operation~\cite{Pozzato.2023, Aitio.2021}. However, capacity remains the primary warranty-relevant SOH metric~\cite{Figgener.2024}. Finally, reproducibility remains restricted by the limited availability of public code and open long-term field datasets with validated reference measurements. Among the reviewed studies, the home energy storage systems (HESS) datasets published by Figgener et al.~\cite{Figgener.2024} represent a notable exception by combining multi-year field operation data with periodic reference capacity tests.
Taken together, the literature indicates that it is still insufficiently explored whether laboratory-trained cell-level degradation models can be transferred successfully to system-level, capacity-based SOH prognosis for real-world BESS operation. Existing studies are only rarely benchmarked against field capacity measurements, and predictive uncertainty is often not quantified. Limitations that prevent simple upscaling from cell-level to system-level arise from two broad classes of uncertainty sources: systematic effects that are reducible in principle, and stochastic effects that are not. System-level operating effects are systematic in nature. They arise from spatially and electrically heterogeneous conditions within the battery system, such as thermal gradients, current imbalance between parallel strings, or variations in position-dependent contact resistance~\cite{Reniers.2023}. These effects are governed by identifiable physical principles and could be modeled given sufficiently detailed knowledge (for instance by using finer system instrumentation). Manufacturing-related cell-to-cell variability in initial capacity and internal resistance falls into the same class, since it is non-destructively measurable prior to deployment~\cite{Baumhofer.2014}. The contribution of systematic effects to cell-to-system discrepancy is therefore reducible, similar to epistemic uncertainty.
A second class comprises the intrinsic stochasticity of electrochemical degradation itself~\cite{Barbers.2024, Beck.2021}. This divergence in individual cells' aging trajectories persists even among cells with comparable initial state. This effect remains even under perfect model knowledge and can only be characterized statistically, similar to aleatoric uncertainty. Consequently, improving system-level predictions requires both improved representation of systematic effects and better statistical characterization of stochastic aging divergence. 

This work addresses this gap by applying a validated probabilistic cell-level degradation model to system-level SOH prognosis for LFP-based residential BESS and benchmarking the resulting predictions against multi-year field capacity measurements. Divergence in aging trajectories, learned from cell-level replicate data, is propagated into every predicted trajectory (section~\ref{sec:uncertainty}), while the cell-to-system approximation (section~\ref{sec:celltosystem}) brackets systematic effects arising from temperature and current heterogeneity. In this way, the study explores how well-established cell-level degradation models can be leveraged for system-level prognosis. Furthermore, the study analyses the extent to which laboratory training data constrains predictive confidence under real field conditions, with resulting implications for future aging test design.

\subsection{Previous Work}
Naumann et al. \cite{Naumann.2018, Naumann.2020} carried out an extensive aging study on commercial LiFePO$_4$/graphite cells (Sony/Murata). This laboratory aging study spans a broad test matrix covering a wide range of stress factors: 17 test conditions for calendar aging and 19 test conditions for cyclic aging for a duration of 885 days. Each test condition was applied to three cells at the same time, enabling estimation of aleatoric uncertainty. 
Furthermore, Naumann et al. presented a semi-empirical aging model for the LiFePO$_4$/graphite cell, which is part of the time-series simulation tool SimSES for energy storage systems~\cite{Moller.2022}. Gasper et al. developed a ML-assisted bootstrapped degradation model based on the same dataset~\cite{Gasper.2022, Collath.2021}.  
A previous work~\cite{Graner.2026} established the open-source probabilistic capacity-loss prediction framework based on stochastic degradation trajectories using the LFP cell aging datasets by Naumann et al.~\cite{Naumann.2018, Naumann.2020}. Details on data preprocessing, feature extraction, model architecture, and training procedures are documented in the open-source repository~\cite{BattProDeep.github}. The present work extends that framework with quantified cell-level validation accuracy, a systematic uncertainty-driver analysis comparing laboratory and field stress-factor distributions, and design recommendations for future laboratory aging campaigns.
The field datasets published by Figgener et al.~\cite{Figgener.2024} are particularly suitable for application to our framework, because a subset of the investigated HESS uses the same commercial LiFePO$_4$/graphite cell type as the laboratory aging study. This creates a rare opportunity to benchmark laboratory-trained probabilistic aging models against multi-year field operation data and periodic system-level capacity measurements within a consistent chemistry and application domain.

\section{Methodology}
\label{sec:framework}
\begin{figure*}[tb]
    \centering
    \includegraphics[width=1\textwidth]{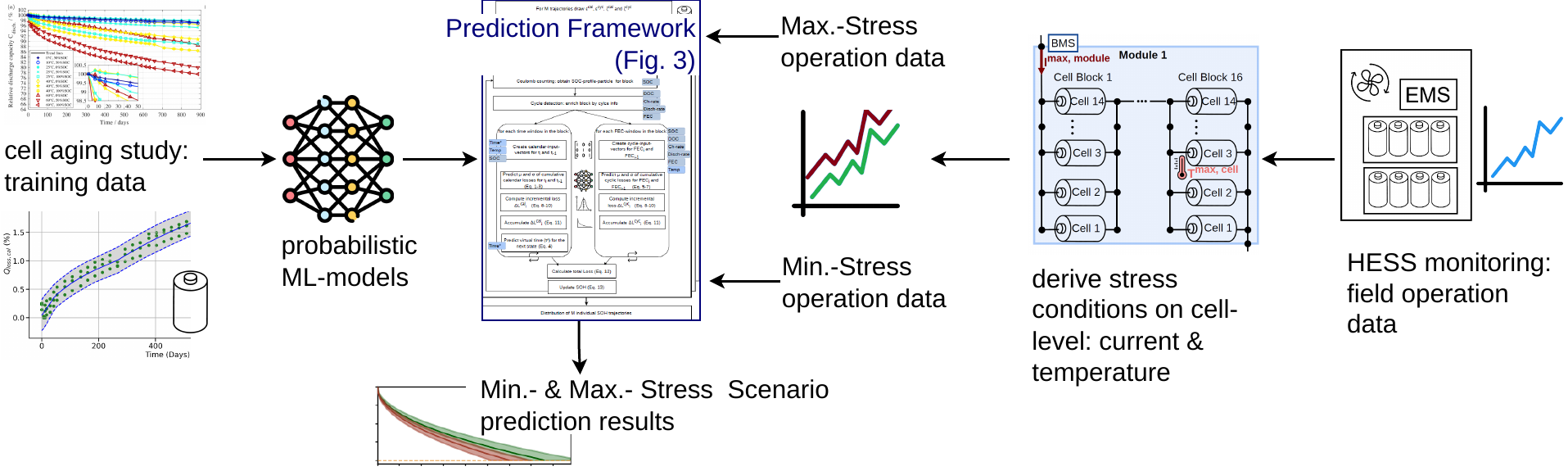} 
    \caption{Illustration of the overall methodology and framework usage. Cell-level aging study data for a LFP battery by Naumann et al.~\cite{Naumann.2018, Naumann.2020} (left) is used to train the probabilistic ML models. Field usage data derived from monitoring of residential energy storage systems (right) is processed to extract cell-level current and temperature, from which the Minimum- and Maximum-Stress Scenario datasets are derived, bounding the range within true system-level degradation is expected to fall. Both are passed through the prediction framework detailed in Fig.~\ref{fig:framework}, yielding the 
    Minimum- and Maximum-Stress degradation trajectory predictions.}
    \label{fig:methodology}
\end{figure*}

The probabilistic degradation-modeling pipeline used in this study is implemented and openly maintained in the "BattProDeep" (BPD) Git repository~\cite{BattProDeep.github} and presented in an earlier contribution~\cite{Graner.2026}. The graphical overview in Fig.~\ref{fig:methodology} depicts how probabilistic ML models trained on laboratory aging study datasets of LFP cells are used in the framework described in section~\ref{sec:pred_framework}. 
The uncertainty of predictions is modeled on the trajectory level, representing variations in aging trajectory evolution as described in section~\ref{sec:uncertainty}. 
The framework needs battery cycling data, which can either be provided on cell-level as is the case in our validation showcase in section~\ref{sec:val_cell_level}, or on system-level. To predict degradation on dynamic usage profiles of full systems, the two extreme stress states Minimum- and Maximum-Stress Scenario on cell-level are identified as described in section~\ref{sec:celltosystem} and processed by the prediction framework.
 
\subsection{The Prediction Framework}
\label{sec:pred_framework}
\begin{figure}[tb]
    \centering
    \includegraphics[width=1\columnwidth]{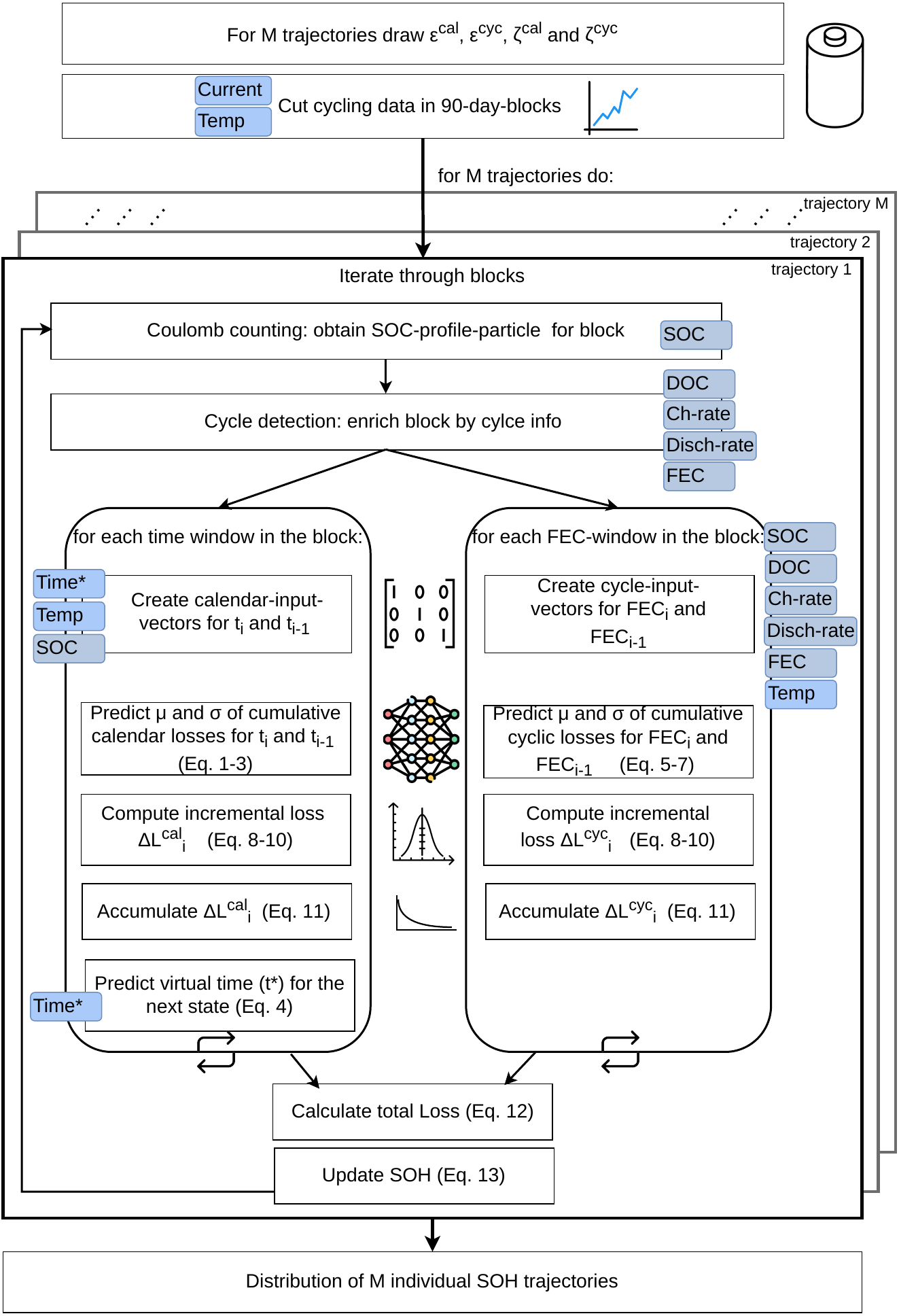} 
    \caption{Overview of the probabilistic ML-based aging prediction framework (BPD-framework), applied in a feedback loop (ensuring the current aging state is accounted for during preprocessing) to predict battery degradation trajectories under dynamic stress conditions. Two separate models estimate the mean and standard deviation of the relative cumulative calendar and cyclic losses, $L^{\mathrm{cal.}}$ and $L^{\mathrm{cyc.}}$, from which the incremental loss $\Delta L$ at each timestep or full equivalent cycle (FEC) is derived and accumulated into the total degradation trajectory. Calendar aging prediction relies on the virtual-time concept $t^*$ of Naumann et al.~\cite{Naumann.2018}.}
    \label{fig:framework}
\end{figure}
The two calendar and cyclic prediction models are trained to predict the relative cumulative capacity loss (relative to the nominal capacity) under static stress conditions as a conditional Gaussian distribution. There is the calendar capacity loss predictor
\[ L^{\mathrm{cal}} \mid \mathbf{x^{\mathrm{cal}}} \sim \mathcal{N}\bigl(f_\mu(\mathbf{x^{\mathrm{cal}}}), f_\sigma(\mathbf{x^{\mathrm{cal}}})\bigr) \] with the vector $\mathbf{x^{\mathrm{cal}}}$ carrying the calendar stress features and there is the cyclic capacity loss predictor \[ L^{\mathrm{cyc}} \mid \mathbf{x^{\mathrm{cyc}}} \sim \mathcal{N}\bigl(g_\mu(\mathbf{x^{\mathrm{cyc}}}), g_\sigma(\mathbf{x^{\mathrm{cyc}}})\bigr) \] with the vector $\mathbf{x^{\mathrm{cyc}}}$ carrying the cyclic stress features. Both predictors are implemented as feedforward neural networks trained to output the parameters of a Gaussian distribution by minimizing the negative log-likelihood. The relative cumulative capacity loss is linked to key stress factors like temperature, SOC, charge rates, and depth-of-cycle (DOC) in the fashion of a feature-based ML model where stress factors are identified from the time-series cycling data during preprocessing. As feature-based models are structurally simpler than those trained directly on raw input data, their training-data demands are correspondingly lower~\cite{Sulzer.2021}. While the models capture nonlinear degradation trends at beginning of life, the training data covers degradation observations only down to approximately 80\% SOH~\cite{Naumann.2018, Naumann.2020} and therefore does not support late-life knee-point behavior. The two models are applied to predict the mean relative cumulative capacity loss $\mu^{L}$ and its standard deviation $\sigma^{L^{\mathrm{al}}}$. The latter accounts for variability in the training data and therefore describes the aleatoric uncertainty of the prediction. 

Furthermore, epistemic uncertainty regarding the prediction of $\mu^{L}$ itself is reflected by an ensemble of $K$ bootstrap-trained models (similar to the approach in \cite{Gasper.2022}), yielding a separate $\sigma^{L^{\mathrm{ep}}}$. The bootstrap members are obtained by varying the influence of each test condition and their progression along time or FEC in the training datasets across the ensemble members. Thus, the ensemble reflects model-driven uncertainty due to the unevenly distributed training data, limited coverage of operating conditions, and predictions beyond the observed operating regime. 

Laboratory studies record aging under static stress states, for example fixed temperature and fixed storage SOC. To predict capacity loss under dynamic battery operation, the total degradation curve is instead formulated as a cumulative process. It is built from short time-based intervals for calendar aging, or FEC-based intervals for cyclic aging, each contributing an incremental capacity loss according to the stress state during that interval (see Fig.~\ref{fig:inc_loss_vt} in the Appendix~\ref{appendix:cum_loss_virt_time}). The total degradation trajectory will consist of all those incremental loss segments.  
Our prediction framework (further called BPD-framework), illustrated in Fig.~\ref{fig:framework}, partitions the multi-year cycling data into sequential 90-day blocks. Each block is preprocessed independently to identify half-cycles (in both charge and discharge direction) and to derive the features needed for the subsequent prediction step: depth-of-cycle, charge/discharge rates, and number of full equivalent cycles (FEC). The remaining capacity is updated and used to compute the SOC profile of the subsequent block, reflecting the progressive reduction in usable capacity.

%\subsection{Dynamic Calendar Aging Prediction}
%\label{sec:cal_aging}
\medskip
\textbf{Dynamic Calendar Aging Prediction} 

\noindent Feature vectors are generated for the beginning and end of each time window $w$ (by default one hour) in a block. The concept of $virtual~time$ from Naumann et al.~\cite{Naumann.2018} is applied to capture the cumulative effect of capacity loss across the history of varying stress states (see Fig.~\ref{fig:inc_loss_vt} in the Appendix~\ref{appendix:cum_loss_virt_time}). 
%The current $virtual~time$ ($t^*_{i}$) is calculated as $t^*_{i}=t^*_{i-1}+\Delta t$, where ($\Delta t$) is the duration of the window. 
The input vectors for the calendar aging predictor consist of mean SOC, mean temperature (averaged over the timesteps in $w$) and virtual time: $\mathbf{x}_{i}^{\mathrm{cal}} = (t_i^*, \overline{\mathrm{T}}, \overline{\mathrm{SOC}})$. The relative mean total capacity loss $\mu^{L^{\mathrm{cal}}}$ and its standard deviations $\sigma^{L^{\mathrm{cal; al}}}$ and $\sigma^{L^{\mathrm{cal; ep}}}$ are predicted using: 
\begin{align}
\mu_{i}^{L^{\mathrm{cal}}}
    &= f_\mu\!\left(\mathbf{x}_{i}^{\mathrm{cal}}\right), \label{eq:16} \\
\sigma_{i}^{L^{\mathrm{cal;al}}}
    &= f_\sigma\!\left(\mathbf{x}_{i}^{\mathrm{cal}}\right), \label{eq:17} \\
\mu_{i}^{L^{(k)}}
    &= f_\mu^{(k)}\!\left(\mathbf{x}_{i}^{\mathrm{cal}}\right), \qquad \mathrm{with}\ k=1,\dots,K, \notag \\
\sigma_{i}^{L^{\mathrm{cal;ep}}}
    &= \operatorname{Std}\!\left(\mu_{i}^{L^{(1)}}, \dots, \mu_{i}^{L^{(K)}}\right). \label{eq:18}
\end{align}
An inverse virtual~time~model~$f^{-1}$ provides the updated $virtual~time$ for the next prediction step based on the relative cumulative capacity loss $L^{\mathrm{cal}}$.
\begin{equation}
t^*_{i-1} = f^{-1} \bigl(L^{\mathrm{cal}}, \overline{T}, \overline{\mathrm{SOC}} \bigr) \label{eq:21}
\end{equation}
%\subsection{Dynamic Cyclic Aging Predictor}
%\label{sec:cyc_aging}
\medskip
\textbf{Dynamic Cyclic Aging Prediction} 

\noindent
Feature vectors are created for the beginning and end of each FEC window ($w_{fec}$, by default one FEC) in a block. Each vector contains averaged SOC, temperature, charge rate (C-rate), discharge rate (Disch-rate) and maximum DOC across the partial cycles as well as FEC:
\[
\mathbf{x}_i^{\mathrm{cyc}} = \left(\mathrm{FEC}_i,\ \overline{\mathrm{Ch\text{-}rate}},\ \overline{\mathrm{Disch\text{-}rate}},\ \overline{\mathrm{T}},\ \overline{\mathrm{SOC}},\ \mathrm{DOC}\right).
\]
\begin{comment}
The current number of FEC, is updated as $FEC_{i} = FEC_{i-1} + \Delta FEC $, where $\Delta FEC$ is the incremental change in number of FECs identified during preprocessing. The average charge rate ($\overline{\text{C-Rate}}$) and discharge rate ($\overline{\text{Disc-Rate}}$), the average SOC ($\overline{\text{SOC}}$) and average temperature ($\overline{\text{T}}$) are the mean over the number of timesteps within $w_{fec}$. The relevant DOC for $w_{fec}$  is determined by the maximum depth observed during either charging or discharging within that window. Specifically, the DOC is defined as the largest absolute change in SOC during a charge or discharge event:
\begin{align}
DOC = \max_{i \in w_{\mathrm{fec}}} \left(
\mathrm{DepthOfC.}_i,
\mathrm{DepthOfDisc.}_i
\right) \label{eq:28}
\end{align}

\end{comment}
The relative mean total capacity loss $\mu^{L^{\mathrm{cyc}}}$ and its standard deviation $\sigma^{L^{\mathrm{cyc; al}}}$ and $\sigma^{L^{\mathrm{cyc; ep}}}$ are predicted using:
\begin{align}
\mu_{i}^{L^{\mathrm{cyc}}}
    &= g_\mu\!\left(\mathbf{x}_{i}^{\mathrm{cyc}}\right), \label{eq:22} \\
\sigma_{i}^{L^{\mathrm{cyc;al}}}
    &= g_\sigma\!\left(\mathbf{x}_{i}^{\mathrm{cyc}}\right), \label{eq:23} \\
\mu_{i}^{L^{(k)}}
    &= g_\mu^{(k)}\!\left(\mathbf{x}_{i}^{\mathrm{cyc}}\right), \qquad \mathrm{with}\ k=1,\dots,K, \notag \\
\sigma_{i}^{L^{\mathrm{cyc;ep}}}
    &= \operatorname{Std}\!\left(\mu_{i}^{L^{(1)}}, \dots, \mu_{i}^{L^{(K)}}\right). \label{eq:24}
\end{align}

\subsection{Uncertainty Quantification and Stochastic Trajectory Interpretation}
\label{sec:uncertainty}
As described in~\cite{Graner.2026}, uncertainty can be modeled at the trajectory level rather than a sequence of statistically independent per-step fluctuations. To this end, an ensemble of $M$ trajectories is generated, where each trajectory represents one plausible degradation path. For each trajectory $m$, stochastic deviation parameters are obtained via sampling of the standard normal distribution: $\varepsilon^{\mathrm{cal(m)}}$, $\zeta^{\mathrm{cal(m)}}$, $\varepsilon^{\mathrm{cyc(m)}}$ and $\zeta^{\mathrm{cyc (m)}}$. The use of separate stochastic terms for calendar and cyclic aging is consistent with related literature \cite{Paul.2013}. $\varepsilon$ and $\zeta$ are scalar weights drawn once per trajectory and held fixed across all timesteps. Multiplied with the predicted aleatoric and epistemic standard deviations (Eq.~\ref{eq:8} and~\ref{eq:9}), they determine a trajectory's persistent deviation from the mean prediction. The stochastic incremental relative losses for trajectory $m$ are then given by
\begin{align}
L^{{(m)}}_{i-1}
&=
\mu^{L_{i-1}}
+
\zeta^m \cdot \sigma^{L^{\mathrm{ep}}}_{i-1}
+
\varepsilon^m \cdot \sigma^{L^{\mathrm{al}}}_{i-1}, \label{eq:8}\\
L^{{(m)}}_{i}
&=
\mu^{L_{i}}
+
\zeta^m \cdot \sigma^{L^{\mathrm{ep}}}_{i}
+
\varepsilon^m \cdot \sigma^{L^{\mathrm{al}}}_{i},\label{eq:9} \\
\Delta L_i^{{(m)}}
&=
L^{{(m)}}_{i}
-
L^{(m)}_{i-1}
\end{align}
for incremental calendar losses $\Delta L^{\mathrm{cal}(m)}$ and incremental cyclic losses $\Delta L^{\mathrm{cyc}(m)}$ separately.
As negative capacity loss is physically implausible in the context of battery aging, any negative outputs are clipped to zero.
%The first particle is kept on the central trajectory, providing a direct deterministic reference within the ensemble.
%We chose a conservative setting of applying the same $\varepsilon_m$ to both calendar and cyclic aging contributions. The stochastic incremental loss for each particle $m$ is derived from the predicted cumulative mean and standard deviation:
The accumulated relative loss for calendar and cyclic aging, for each trajectory, is then obtained by
\begin{equation}
L_i^{\mathrm{cal, cyc}(m)}
= \sum_{j=1}^{i} \Delta L^{\mathrm{cal, cyc}(m)}_j 
\end{equation}
from which the relative total loss for each trajectory $m$ is derived as
\begin{equation}
L_i^{\mathrm{total}(m)} = L_i^{\mathrm{cal}(m)} + L_i^{\mathrm{cyc}(m)}.
\end{equation}
The SOH trajectories for each trajectory $m$ are derived as
\begin{equation}
\mathrm{SOH}^{(m)}_i
= 1 - L^{\mathrm{total}(m)}_{i}.
\end{equation}
%In summary, the proposed framework provides a physically consistent and statistically coherent representation of uncertainty for cumulative battery degradation under dynamic operating conditions. 
%By operating at the level of stochastic trajectories, it naturally respects the temporal structure of aging while fully utilizing the available training data.

 \begin{comment}   
 
\subsubsection{Key Takeaways}
\begin{itemize}
    \item Virtual time \(t_i^*\) encodes stress history \textit{implicitly}, avoiding explicit dependencies on all past states.
    \item Stochastic trajectories with fixed \(\epsilon_m\) ensure uncertainty is \textit{physically correlated} across timesteps.
    \item This approach avoids artificial uncertainty inflation and aligns with the \textit{cumulative nature} of battery degradation.
\end{itemize}
\end{comment}

\subsection{Scaling from System to Cell-Level}
\label{sec:celltosystem}
Cell-to-cell variances caused by manufacturing variability can lead to inhomogeneous system states in battery packs and to current and voltage imbalance among the subunits. These effects are further amplified by thermal gradients at pack or rack level and tend to increase with lifetime. 
\begin{figure}[htb]
    \centering
    \includegraphics[width=1\columnwidth]{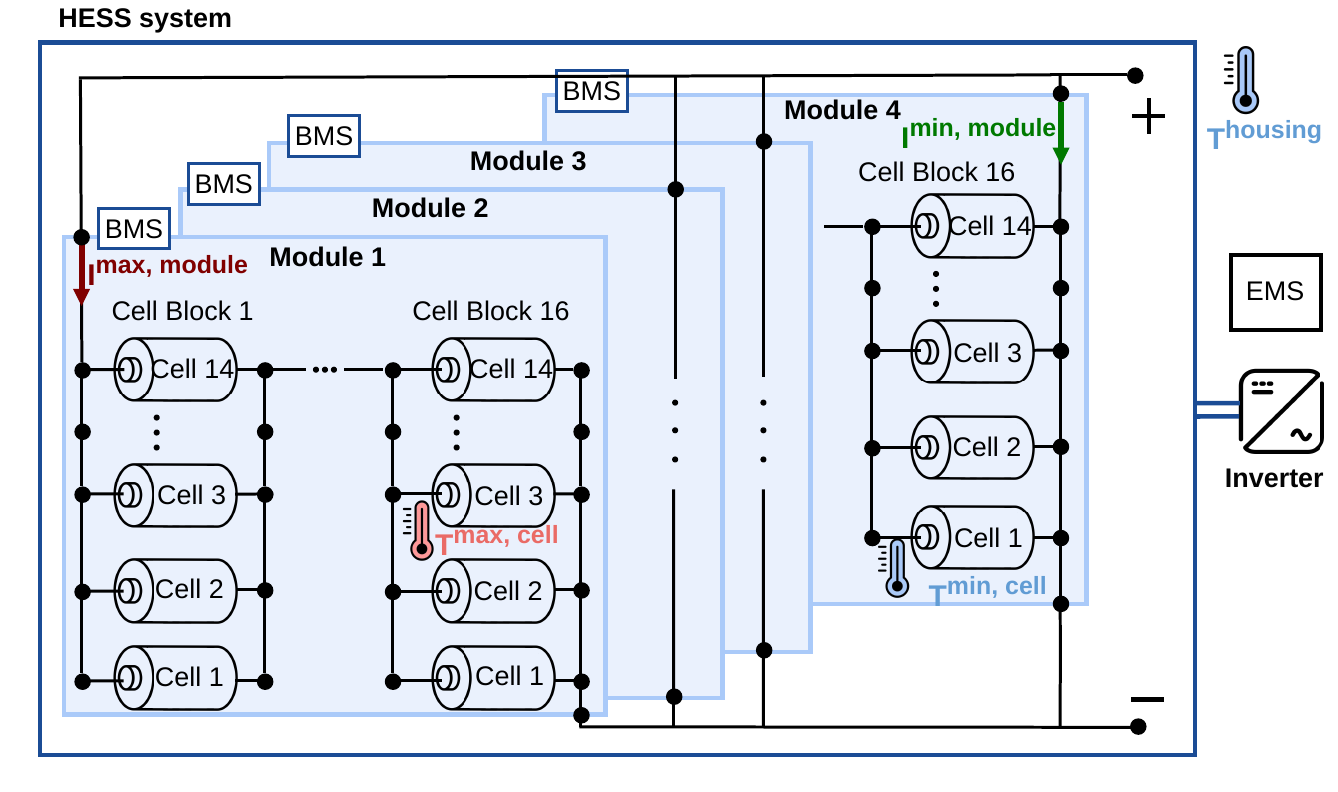} 
    \caption{Exemplary topology of a HESS. The battery pack consists of multiple modules connected in parallel (Module 1 ... 4), each containing single cells in parallel and series connection (16s14p). Battery management systems (BMS) monitor the modules, while an energy management system (EMS) coordinates the system-level charging and discharging decisions. Current and voltage imbalance among the modules and single cells may lead to heterogeneous charging states and stress conditions. Inhomogeneous thermal propagation may amplify the latter. The figure exemplarily shows that a cell can be exposed to both higher currents and temperatures ($T^{max}$ and $I^{max}$) than a cell that faces minimal stress ($T^{min}$ and $I^{min}$). In the long run this might result in inhomogeneous aging states of single cells.}
    \label{fig:bess_topology}
\end{figure}
Fig.~\ref{fig:bess_topology} depicts how current and temperature across individual modules and cells can vary. Two limiting cell-level stress conditions provide a physically motivated approximation for plausible system-level degradation in this study: a Maximum-Stress Scenario, combining the highest recorded module current and cell temperature, and a Minimum-Stress Scenario, combining the lowest recorded values of both. 
The Maximum-Stress Scenario causes the highest capacity loss for individual cells, making it particularly relevant for assessing system-level degradation. This is because the weakest subunit in a series connection limits a string's capacity and ultimately lowers the overall SOH of the system \cite{Bulow.2023}.

\section{Results}
\label{sec:results}
This section describes first the validation of the probabilistic battery degradation framework introduced in section~\ref{sec:framework} on cell-level: in section~\ref{sec:val_cell_level} predictions are made on cell's cycling data, that represent typical BESS-use cases. Thanks to granular check-up measurements, these predictions can be evaluated regarding their accuracy. Section~\ref{sec:case_study} then introduces the application of the BPD-framework to field system operation data. 

\subsection{Validation on Cell-Level}
\label{sec:val_cell_level}
Two validation profiles are used to represent typical usage scenarios of classical battery energy storage systems: a  residential photovoltaic (PV) home energy storage system (PV-HESS) for self-consumption of generated PV power and a second PV-system with additional provision of frequency containment reserve power (PV-FCR)~\cite{Naumann.2020}. 
The two load profiles were used by Naumann et al.~\cite{Naumann.2020} to repeatedly cycle multiple LiFePO$_4$/graphite cells in the laboratory, performing regular check-up measurements to obtain validation datasets for the various degradation models developed for this cell~\cite{Naumann.2018, Naumann.2020, Gasper.2022}. These profiles exhibit distinct characteristics. In the PV-FCR profile, the SOC varies between 36.3\% and 61.4\% and the cells are cycled with low charge and discharge rates (maximum 0.36 C) at a constant temperature of $40.6\,^\circ\mathrm{C}$.
In the PV-HESS profile, the SOC-range is higher than in the PV-FCR profile (from 5.4\% to 80.0\%), as are the charge and discharge rates (maximum 0.75 C). The temperature is constant at $40.8\,^\circ\mathrm{C}$. For both profiles, the measured nominal battery cell capacity at the beginning of testing was 3 Ah. 
\begin{figure*}[hbt]
    \centering
    \includegraphics[width=1\textwidth]{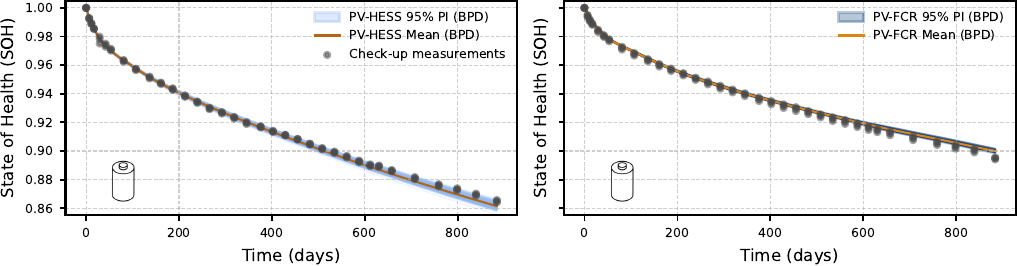} 
    \caption{Results for the model validation on the dynamic cell aging test profiles representing (1) a PV home energy storage system use case (PV-HESS) and (2) a PV-system use case with additional provision of frequency containment reserve power (PV-FCR). The BattProDeep-framework introduced in the earlier sections computes multiple possible aging trajectories from which the prediction intervals and mean aging behavior are derived (indicated as "BPD" in the legend). The SOH derived from the capacity check-up measurements aligns well with the trajectories predicted.}
    \label{fig:validation_data}
\end{figure*}

As described in section~\ref{sec:uncertainty}, uncertainty is represented by stochastic degradation trajectories. %For each input profile, we sample 100 trajectories using the BPD-framework. %Because different incremental losses are predicted for each particle, the resulting degradation trajectories diverge over time. This divergence also propagates to the preprocessing stage: different SOC profiles are obtained for different remaining capacities during coulomb counting, as well as different virtual times for the cumulative loss states, leading to distinct input feature vectors for each particle. 
Fig.~\ref{fig:validation_data} displays the mean of all trajectories as well as the 95\% prediction interval (PI).
The mean trajectories of both profiles show the characteristic square-root dependent drop in SOH. They show final SOH levels of 86.1\% for the PV-HESS profile and 90.0\% for the PV-FCR profile. Most of the capacity check-up measurement points lie close to or within the PIs, while only the last check-up measurement for the PV-FCR profile lies outside these bounds. The PV-HESS profile shows a wider PI compared to the PV-FCR validation profile, which will be discussed in section~\ref{sec:uncertainty_drivers}. Compared with previously developed probabilistic prediction models for this specific LFP dataset~\cite{Gasper.2022}, the BPD-framework shows improved prediction accuracy. Measured for the trajectory mean, BPD achieves a mean absolute SOH prediction error (MAE) of approximately 0.4\%, compared to a MAE of approximately 0.9\% for the bootstrapped semi-empirical degradation model~\cite{Gasper.2022}) in both profiles. A possible explanation is that the parameter submodels in~\cite{Gasper.2022} are fit to values that are first reduced to a single scalar per test condition and then generalized across conditions. BPD instead trains directly on the full time-resolved data, without this per-condition reduction or generalization. This methodological difference may explain the gap in accuracy.

\subsection{Benchmarking Case Study on Field Systems}
\label{sec:case_study}
In this section, we apply the probabilistic battery degradation framework introduced in section~\ref{sec:framework} to field system operation data acquired by Figgener et al. in~\cite{Figgener.2024} in order to predict system-level degradation behavior. Capacity test measurements available alongside the field data are used for benchmarking the results.

\subsubsection{Data}
\label{sec:data}
The datasets published by Figgener et al.~\cite{Figgener.2024} comprise multi-year field measurements of 21 private HESS combined with residential photovoltaic installations, capturing current, voltage, power on system level, as well as pack housing temperature and room temperature. The case study focuses on one representative system (system 14), as it uses the same commercial LiFePO$_4$/graphite cells the ML models are trained on. For this system, we additionally hold a dataset of monitoring data from the HESS manufacturer ourselves, including more granular field-operation data (called BPD-dataset in the following sections). In addition to system-level data, it contains minimum and maximum current recorded on module level, indicating the extrema of current imbalance between modules. Additionally, it contains the recorded minimum and maximum temperature of all sensors on cell-level, indicating the temperature spread across the system. Table~\ref{tab:dataset_properties} shows the relevant variables contained in both datasets. 

\begin{table}[htb]
    \centering
    \caption{Monitored variables available in the Figgener et al.\ and BPD-datasets for the BESS system in field operation.}
    \label{tab:dataset_properties}
    \begin{tabular}{lll}
        \hline
        Variable & Figgener et al.~\cite{Figgener.2024} & BPD-dataset \\
        \hline
        Current      &System current & System current \\
             & --         & Min./max.  \\
             &         & module current\\
        Temperature    &Housing         &Min./max. \\
           &  temperature        & cell temperature\\
        Capacity measurem.    &Three in field       & One at BOL  \\
        \hline
    \end{tabular}
\end{table}
%Again it has to be noted, that it is not indicated, which sensor (means which area of cells) recorded the minimum or maximum temperature value. 

\begin{figure}[htb]
    \centering
    \includegraphics[width=1\columnwidth]{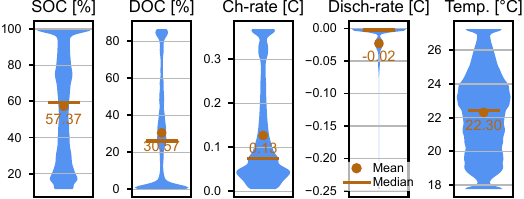} 
    \caption{Battery stress factor distributions derived from field operation of the relevant LFP systems in~\cite{Figgener.2024}}. 
    \label{fig:field_violins}
\end{figure}

The HESS operation shows pronounced daily and seasonal patterns. Charging generally occurs at higher power during daytime PV generation, while discharge predominantly takes place at night at lower rates to meet household standby consumption (see Fig.~\ref{fig:energy_throughput} in the Appendix showing average hourly energy throughput (Wh) of a HESS throughout a typical day). Overall, the systems spend approximately 48.5\% of the time discharging, 34.7\% in idle mode, and 16.8\% charging. Over a year, on average 152 FECs occur. Fig.~\ref{fig:field_violins} visualizes the battery stress factors derived from the usage data. The bimodal SOC distribution reflects typical HESS operation where batteries are often fully charged during daylight hours and rest at low SOCs overnight. The cycle depth is mostly below 10\%, indicating shallow cycles, with occasional deeper cycles reaching up to 85\%. Charge rates are generally low with a mean of approximately 0.13 C. Discharge rates are even lower with a mean of about 0.02 C. The storage housing temperatures range mostly from $18\,^\circ\mathrm{C}$ to $28\,^\circ\mathrm{C}$ with an approximate mean of $22.3\,^\circ\mathrm{C}$. %This goes well in line with the optimal operating temperature range for BESS of 15–35 °C indicated in literature~\cite{Ma.2018, Gungor.2023}. 

Seasonal PV output in Germany is strongly skewed toward summer (roughly 70\% of annual generation~\cite{Figgener.2024}), which is observable in the stress factors' seasonal patterns: extended periods at high SOC and increased temperatures during summer, as systems spend many hours at full charge (see Fig.~\ref{fig:field_violins_seasonal} in the Appendix~\ref{appendix:operational_data} for more details).
 
%Maximum system power is constrained by the inverter and battery design, and even identical systems exhibit variability due to differences in PV generation and household load. 
 
%This reflects typical storage usage, which mostly consists of full charging and discharging interspersed with long resting times.  

%The characteristics are largely determined by system design, particularly the ratio of battery capacity to inverter power, which governs achievable C-rates and cycling frequency. 
This field data characterization provides context for the operational environment in which our framework is applied. The recurring daily and seasonal structures furthermore enable extrapolation of HESS operation and support degradation prediction in the future.
%\textcolor{blue}{the systems show different operational strategies (Supplementary Fig. 1) }

\subsubsection{Obtaining Cell-Level Data}
\label{sec:fieldtosystem}
%(168 Ah system capacity). 
%Figure~\ref{fig:bess_topology} illustrates the system topology of $system~14$ that is considered for scaling the operation data down to cell-level and determining the time-series for the Maximum- and Minimum Stress Scenario. 
As depicted in Fig.~\ref{fig:methodology}, system-level degradation is approximated by the degradation trajectories of the Minimum- and Maximum-Stress Scenario on cell-level. As cell-level currents are not measured, we estimate cell-level currents $I_i^{min,cell}$ and $I_i^{max,cell}$ by applying Kirchhoff's law while neglecting cell-individual resistance spread. The minimum ($I_i^{min,module}$) and maximum module currents ($I_i^{max,module}$) provided in the BPD-dataset are therefore divided by the number of parallel-connected cells within each module using Eq.~\ref{eq:current_split} and assuming equal current split within the modules for simplification. As drafted in Fig.~\ref{fig:bess_topology}, each module consists of 16 series-connected cell blocks consisting of 14 parallel cells (16s14p). 
\begin{equation}
    I_i^{min/max,cell} = \frac{I_i^{min/max,module}}{n^{parallel}} \label{eq:current_split}
\end{equation}
%\[\]
For temperature, the dataset of Figgener et al. includes pack housing temperature and room temperature~\cite{Figgener.2024}. The pack housing temperature does not reflect the actual battery temperature (cell-surface), which is why for the simulation case study we use the temperature recordings $T_i^{\mathrm{max,cell}}$ and $T_i^{\mathrm{min,cell}}$ available in the BPD-dataset. These were obtained from sensors attached to the battery cells. Their difference is quite small with $1.56\,^\circ\mathrm{C}$ on average.   
By combining the minimum and maximum values for current and temperature on cell-level, we obtain the Minimum- and Maximum-Stress Scenarios as described in section~\ref{sec:celltosystem}.

\subsubsection{Prediction Results and Evaluation}
\label{sec:cs_results}
\begin{figure*}[tb]
  \centering
  \includegraphics[width=\textwidth]{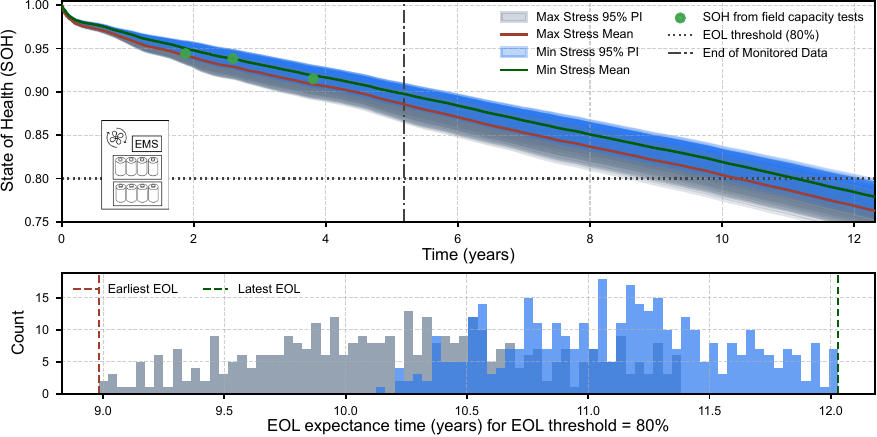}
  \caption{Benchmark of the prediction framework against SOH values derived from field capacity measurements. (Top) Predicted SOH evolution for the Minimum- and Maximum-Stress Scenarios, with the 95\% prediction interval (PI) shown as a shaded band. The SOH values derived from field capacity measurements coincide well with the 95\% PIs for stress factor derived and topology scaled system predictions. (Bottom) Distribution of expected EOL times for both scenarios.}  
  \label{fig:Figgener14results}
\end{figure*}
This section presents the system-level degradation predictions for the Minimum- and Maximum-Stress Scenarios of one exemplary LFP HESS, together with their corresponding 95\% PIs.  For each scenario, approximately 400 trajectories are simulated, each representing one plausible degradation path. Figure~\ref{fig:Figgener14results} shows the resulting mean SOH trajectories and empirical 95\% PIs, obtained from the 2.5th and 97.5th percentiles of the trajectory ensemble:
\[\mathrm{PI}_i = \big[\mathrm{SOH}_i^{(2.5\%)},\ \mathrm{SOH}_i^{(97.5\%)}\big].\]
This choice retains the dominant spread of the trajectory ensemble while excluding only the most extreme 5\% of outcomes. The mean SOH trajectories of both scenarios diverge by 1.6\% after 12.3 years.

The original dataset provides 5.2 years of cycling data, which is extrapolated to predict degradation down to the EOL threshold of 80\% SOH. The figure shows where the forecast of future operation begins. Predictions beyond this EOL threshold are not covered by the ML model's training data from the laboratory aging study, which only extends to approximately 80\% SOH. As a result, late-life knee-point behavior cannot be predicted. Figgener et al. performed field capacity tests following the test-protocol described in the supplementary information of~\cite{Figgener.2024}. The resulting values, shown as green markers in Fig.~\ref{fig:Figgener14results}, represent the system's remaining extractable capacity and must therefore be converted to SOH with care. Because the system does not permit discharge below 5\% SOC, the accessible SOC range is limited to 95\%. In addition, the datasheet capacity of 158 Ah differs from the physically available begin-of-life (BOL) capacity on system level. We therefore convert the field capacity measurements to SOH as
\[\mathrm{SOH} = \frac{C_{\mathrm{field}}}{\Delta \mathrm{SOC}\ \cdot C_{\mathrm{BOL}}},\]
where $C_{\mathrm{field}}$ is the extractable field capacity measured in the field capacity tests, $\Delta \mathrm{SOC}$ is the usable SOC range, and $C_{\mathrm{BOL}}$ is the begin-of-life system capacity (which is proprietary to the manufacturer).
All three converted field SOH values (marked as green dots in Fig.~\ref{fig:Figgener14results}) lie within the predicted intervals. The first measurement lies close to the mean trajectory of the Maximum-Stress Scenario, whereas the second measurement lies close to the Minimum-Stress Scenario and the third measurement lies right between both mean trajectories. Overall, the measured system-level SOH is therefore plausibly bracketed by the two cell-level extreme-stress scenarios.

Figure~\ref{fig:Figgener14results} in addition to the SOH trajectories shows the distribution of predicted EOL dates across all simulated paths. The earliest expected trajectory crossing of the 80\% SOH threshold occurs after 9 years for the Maximum-Stress Scenario, whereas the latest threshold-crossing is expected after 12 years for the Minimum-Stress Scenario. The resulting spread of approximately 3 years corresponds to roughly one third of the mean predicted system lifetime. This spread highlights that even moderate differences in long-term degradation behavior can translate into substantial variation in expected system lifetime. The Minimum- and Maximum-Stress Scenarios show only minor differences in operational stress: the temperature levels spread by roughly 5\% and the currents show a spread of 9\%. These small differences in operational stress translate into a difference of 10 months of expected lifetime for the mean degradation trajectories of both scenarios. 

The magnitude of prediction uncertainty in Fig.~\ref{fig:Figgener14results} corresponds to the spread of SOH-estimates obtained by the predictions of Figgener et al. in~\cite{Figgener.2024}. The high spread in lifetime expectancy means manufacturers and operators need to carefully weigh how much risk they want to take, when it comes to maintenance planning or warranty assessment.
%Rather than yielding a single deterministic end-of-life date, the framework provides a probabilistic range of plausible EOL outcomes that may support risk-aware maintenance planning, warranty assessment, and second-life decision-making.

\section{Uncertainty Driver Analysis}
\label{sec:uncertainty_drivers}
The PIs obtained for the SOH trajectory prediction of the field HESS are substantially wider than those of the PV-HESS validation profile, although both represent the same general application domain. In the following, we discuss the wider field prediction intervals through two complementary perspectives: (1) the limited laboratory support for relevant stress conditions, which is likely to increase epistemic uncertainty, and (2) the empirical spread of degradation observations under those conditions, which is likely to increase aleatoric uncertainty. A similar mismatch between laboratory aging studies and field-relevant battery stress in EV usage has also been highlighted in the review by Schreiber et al.~\cite{Schreiber.2026}.

\subsection{Epistemic Uncertainty: Laboratory-Field Alignment}
\label{sec:epistemic_mismatch}
To interpret epistemic uncertainty diagnostically, we introduce three alignment metrics that describe how well the laboratory aging study supports predictions for the field and validation scenarios. These metrics are not a direct measure of epistemic uncertainty itself. Instead, they serve as interpretable proxies for situations in which epistemic uncertainty is expected to increase, for example due to extrapolation beyond the laboratory domain or interpolation in only weakly supported regions.
For a given stress feature, let $\mathcal{X}_{\mathrm{lab}} = \{x_1, \dots, x_J\}$ denote the set of laboratory test points and let $f_i$, with $i = 1, \dots, N_f$, denote the corresponding field or validation samples. We further define the tolerated laboratory range as $\mathcal{I}_{\mathrm{lab}} = [x_{\min}-\epsilon,\; x_{\max}+\epsilon]$, where $x_{\min}$ and $x_{\max}$ denote the minimum and maximum laboratory observations and $\epsilon$ is a feature-specific tolerance ($1\,^\circ\mathrm{C}$ for temperature, 5\% for SOC and DOC and 0.01~C for Ch-~and Disch-rate). In addition, the distance of sample $f_i$ to the nearest laboratory support point is defined as $d_i = \min_{x_j \in \mathcal{X}_{\mathrm{lab}}} |f_i - x_j|$.

\noindent\textbf{Coverage} measures the fraction of field samples that lie inside the overall laboratory value range:
\[\mathrm{Coverage}=\frac{1}{N_f}\sum_{i=1}^{N_f}\mathbf{1}(f_i \in \mathcal{I}_{\mathrm{lab}}).\]
Low Coverage indicates that extrapolation beyond the observed laboratory envelope is required.

\noindent\textbf{Focus Match} measures the fraction of field samples that lie within a tolerance band around any laboratory test point:
\[\mathrm{FocusMatch}=\frac{1}{N_f}\sum_{i=1}^{N_f}\mathbf{1}(d_i \le \epsilon).\]
Thus, even when Coverage is high, Focus Match can remain low if the field data fall into sparsely sampled regions between laboratory conditions.

\noindent\textbf{Nearest Overlap} measures the average proximity to the nearest laboratory support point, normalized by a feature-specific reference scale~$\Delta_{\mathrm{ref}}$ ($10\,^\circ\mathrm{C}$ for temperature, 10\% for SOC and DOC and 0.01~C for Ch- and Disch-rate):
\[\mathrm{NearestOverlap}=\frac{1}{N_f}\sum_{i=1}^{N_f}\max\!\left(0,\;1-\frac{d_i}{\Delta_{\mathrm{ref}}}\right).\]
Low values indicate weak local support even if the field data remain within the nominal laboratory range.

Taken together, these metrics help distinguish between unsupported extrapolation and interpolation within the nominal laboratory range but away from densely sampled test points. High Coverage alone therefore does not necessarily imply strong local support.

\begin{figure}[htb]
    \centering
    \includegraphics[width=1\columnwidth]{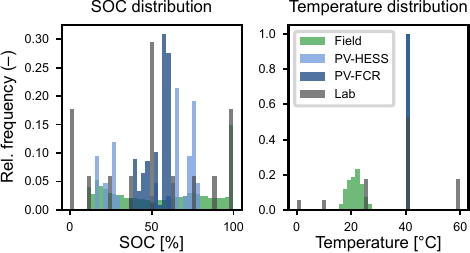} 
    \caption{Comparison of field data and laboratory aging study~\cite{Naumann.2018} for calendar aging stress features. The panels show normalized histograms for the field datasets (Field, PV-HESS, PV-FCR) alongside the laboratory reference (“Lab”), illustrating the distribution overlap. Colors indicate the different datasets.}
    \label{fig:distributions_calendar}
\end{figure}

Figure~\ref{fig:distributions_calendar} compares the stress-factor distributions of the calendar aging study with those of the field and validation datasets, while Table~\ref{tab:calendar_metrics} summarizes the corresponding alignment metrics. Calendar aging studies are particularly difficult to design because significant degradation must be generated within limited experimental time. In practice, this often motivates accelerated testing at elevated temperatures and selected average SOC levels, which need not coincide with stress factors dominating field operation.

For calendar aging, the laboratory support is generally strong for SOC but weaker for temperature. Calendar SOC is well aligned across all datasets, with complete Coverage and high local-support metrics, indicating that average SOC is not a major limitation for the calendar model. Temperature shows a clearer separation: although all datasets remain within the global laboratory range, the field case is much less aligned with the main laboratory test points than the validation profiles. Thus, field temperatures are covered globally but only weakly supported locally. This plausibly contributes to the wider field prediction intervals.

\begin{table}[ht]
\centering
\caption{Calendar aging feature alignment between field datasets and laboratory aging study}
\label{tab:calendar_metrics}
\begin{tabular}{llccc}
\hline
Dataset & Feature & Cover- & Focus & Nearest      \\
 &  & age & Match & Overlap    \\
\hline
Field & SOC & 1.0 & 0.83 & 0.72     \\
PV-HESS & SOC & 1.0 & 0.98 & 0.77    \\
PV-FCR & SOC & 1.0 & 0.93 & 0.67   \\
Field & Temp. & 1.0 & 0.11 & 0.57   \\
PV-HESS & Temp. & 1.0 & 1.00 & 0.92     \\
PV-FCR & Temp. & 1.0 & 1.00 & 0.94    \\
\hline
\end{tabular}
\end{table}

\begin{figure}[htb]
    \centering
    \includegraphics[width=1\columnwidth]{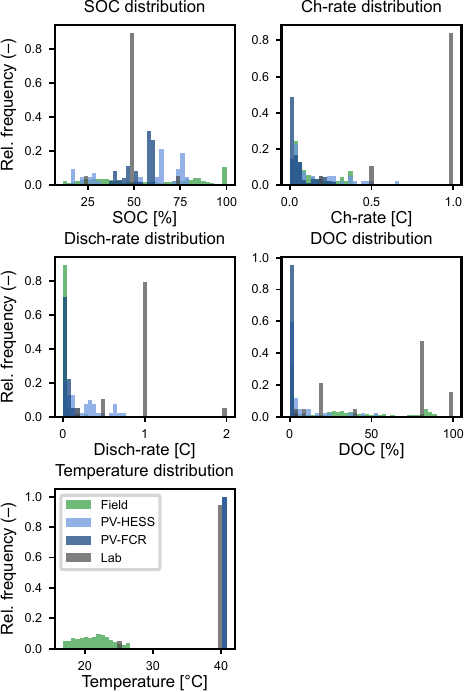 } 
    \caption{Comparison of field data and laboratory aging study~\cite{Naumann.2020} for cyclic aging stress features. The panels show normalized histograms for the field datasets (Field, PV-HESS, PV-FCR) alongside the laboratory reference (“Lab”).}
    \label{fig:distributions_cycle}
\end{figure}

The cyclic aging comparison reveals a much stronger laboratory-field mismatch than the calendar case (Table~\ref{tab:cycle_metrics}, Fig.~\ref{fig:distributions_cycle}). The most severe limitation concerns charge and discharge rates: support is weak for the field data and remains poor even for the validation profiles, especially for discharge rate. This indicates a structural limitation of the cyclic aging study for representing real-world low-rate operation. Cyclic SOC is also less supported than in the calendar case, particularly for the field dataset. DOC, by contrast, remains comparatively well-represented, with complete Coverage for all datasets and only weaker local support in the field scenario. Temperature again sharply separates field and validation data: the field case shows both low range coverage and weak local support, while both validation profiles remain closely aligned with the laboratory conditions.
\begin{table}[ht]
\centering
\caption{Cyclic aging feature alignment between field datasets and laboratory aging study}
\label{tab:cycle_metrics}
\begin{tabular}{llccc}
\hline
Dataset & Feature & Cover- & Focus & Nearest     \\
 &                  & age & Match & Overlap   \\
\hline
Field & SOC & 0.64 & 0.33 & 0.11   \\
PV-HESS & SOC & 0.90 & 0.62 & 0.49    \\
PV-FCR & SOC & 1.00 & 0.23 & 0.26     \\
Field & Ch-rate & 0.26 & 0.02 & 0.00    \\
PV-HESS & Ch-rate & 0.35 & 0.05 & 0.00    \\
PV-FCR & Ch-rate & 0.10 & 0.03 & 0.00     \\
Field & Disch-rate & 0.04 & 0.003 & 0.00     \\
PV-HESS & Disch-rate & 0.49 & 0.03 & 0.00    \\
PV-FCR & Disch-rate & 0.01 & 0.01 & 0.00     \\
Field & DOC & 1.00 & 0.67 & 0.47     \\
PV-HESS & DOC & 1.00 & 0.95 & 0.63     \\
PV-FCR & DOC & 1.00 & 1.00 & 0.55     \\
Field & Temp. & 0.15 & 0.11 & 0.48   \\
PV-HESS & Temp. & 1.00 & 1.00 & 0.92  \\
PV-FCR & Temp. & 1.00 & 1.00 & 0.94    \\
\hline
\end{tabular}
\end{table}

Taken together, the alignment metrics show that the stress conditions in the validation profiles match well with the test-points sampled in the laboratory aging study. In contrast, the field usage scenario more frequently occupies operating states that are weakly supported or unsupported by the laboratory aging study. This provides a plausible explanation for the wider prediction intervals observed for the field predictions, whereas the predictions from the validation dataset show less spread.

Beyond the support mismatches captured by the alignment metrics, real-world BESS usage extends over longer horizons than the laboratory study. The increased uncertainty observed in field predictions is therefore also consistent with long-horizon extrapolation beyond the temporal scope of the laboratory study. Moreover, the field operation involves dynamic usage sequences, including idle phases during which internal gradients may equilibrate. This may reduce degradation severity \cite{Geslin.2024,Schreiber.2025} and may permit partial recovery effects \cite{Schreiber.2025,Lewerenz.2019}. These effects are not isolated by the current alignment analysis, but remain plausible contributors to the epistemic uncertainty observed in field predictions.

\subsection{Aleatoric Uncertainty: Capacity Measurement Spreads}
\label{sec:aleatoric_spread}
Even in well-supported regions of the laboratory design space, predictions remain affected by aleatoric uncertainty, i.e., by intrinsic variability in degradation observations under nominally identical stress conditions. Since calendar aging dominates the predicted long-term degradation for the considered use case, we examine the empirical spread of representative laboratory calendar-aging test points associated with the dominant operating states of each dataset. Larger spreads at these points are expected to correspond to a stronger aleatoric contribution and to wider predicted intervals, provided these states are visited frequently.

The validation datasets (PV-HESS and PV-FCR) are concentrated at temperatures around $40\,^\circ\mathrm{C}$ and within moderate SOC windows below approximately 80\% SOC. As summarized in Table~\ref{tab:spreads}, the corresponding laboratory test points show mostly low relative spreads in measured capacity that translate into observable relative capacity loss spreads, generally below 1\%. The only notable exception is the $40\,^\circ\mathrm{C}$, 0\% SOC test point, where the observable loss spread reaches 2.3\%. Since low-SOC states occur mainly in the PV-HESS profile, this elevated variability is consistent with the slightly wider prediction interval observed for PV-HESS than for PV-FCR (see Fig.~\ref{fig:validation_data}).

\begin{figure}[tb]
  \centering
  \includegraphics[width=\columnwidth]{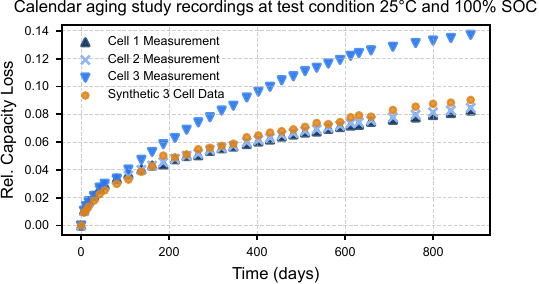}
  \caption{The training data values for relative capacity loss evolution of the three cells tested in the condition $25\,^\circ\mathrm{C}$ and 100\% SOC displayed in shades of blue. Cell 1's capacity loss, markedly higher than the other two replicates, drives the elevated spread and variability in predictions relying on this test point. The orange "Synthetic Cell 1 Data" is relevant for the sensitivity analysis carried out in the following section~\ref{sec:sensitivity_analysis}.}  
  \label{fig:outlier_calendar_point}
\end{figure}

The field dataset is characterized by lower temperatures and extended dwelling at extreme SOC levels. The corresponding representative calendar conditions are therefore centered around $25\,^\circ\mathrm{C}$ and include low, medium, and high SOC states. While the low- and medium-SOC points show only small spreads in observable loss, the high-SOC point at $25\,^\circ\mathrm{C}$ and 100\% SOC exhibits a markedly larger spread in observable relative capacity loss of 5.9\% driven by one cell's drastic capacity loss (see Fig.~\ref{fig:outlier_calendar_point}), exceeding all other considered calendar test points. This indicates that the dominant field calendar states are associated with substantially higher intrinsic degradation variability than the moderate-SOC states that dominate the validation profiles.

%Operational heterogeneity further amplifies aleatoric effects. 
%The validation profiles follow repetitive load trajectories with limited variability, whereas the field dataset aggregates diverse operating modes, including variable temperatures, longer idle periods, and more irregular load amplitudes. The increased variability in stress sequencing introduces additional outcome dispersion even when individual states overlap with laboratory conditions.

\begin{table}[h]
    \centering
    \caption{Relative spread of relevant calendar aging study test points}
    \label{tab:spreads}
    \begin{tabular}{p{0.23\linewidth} p{0.38\linewidth} p{0.23\linewidth}}
        \toprule
        \textbf{Dataset}        & \textbf{Testpoint}        & \textbf{Max. relative spread of loss}                          \\
        \midrule
        Field            & $25\,^\circ\mathrm{C}$ \& 0\% SOC  ,   & 0.3\%\\
        & $25\,^\circ\mathrm{C}$ \& 50\% SOC, & 0.1\% \\
        &$25\,^\circ\mathrm{C}$ \& 100\% SOC& 5.9\%\\
        PV-HESS   &$40\,^\circ\mathrm{C}$ \& 0\% SOC &2.3\%\\        
        &$40\,^\circ\mathrm{C}$ \& 12.5\% SOC &0.1\%\\
        & $40\,^\circ\mathrm{C}$ \& 25\% SOC,    & 0.2\%  \\
        & $40\,^\circ\mathrm{C}$ \& 37.5\% SOC,& 0.4\%\\
        & $40\,^\circ\mathrm{C}$ \& 50\% SOC, & 0.3\%\\
        &$40\,^\circ\mathrm{C}$ \& 62.5\% SOC, & 0.2\%\\
        & $40\,^\circ\mathrm{C}$ \& 75\% SOC & 0.8\% \\
        & $40\,^\circ\mathrm{C}$ \& 87.5\% SOC & 0.2\% \\
        PV-FCR        & $40\,^\circ\mathrm{C}$ \& 37.5\% SOC,  & 0.4\%  \\
        & $40\,^\circ\mathrm{C}$ \& 50\% SOC, &0.3\%\\
        &$40\,^\circ\mathrm{C}$ \& 62.5\% SOC, & 0.2\%\\
        \bottomrule
    \end{tabular}
\end{table}

%Taken together, the analyses of lab-field alignment and capacity measurement spreads provide a coherent explanation for the wider prediction intervals of the field scenario. The field dataset not only occupies operating regions with weaker laboratory support, which is consistent with increased epistemic uncertainty, but also spends substantial time in dominant calendar-aging states, in particular high-SOC regions, that show larger variability in measured degradation. In contrast, the validation profiles remain concentrated in densely sampled laboratory regions with comparatively small observed spread. The wider field prediction intervals are therefore consistent with the combined effect of weaker laboratory support and a plausibly stronger aleatoric contribution under real-world operation.

\subsection{Uncertainty Sensitivity Analysis}
\label{sec:sensitivity_analysis}
\begin{figure}[tb]
  \centering
  \includegraphics[width=\columnwidth]{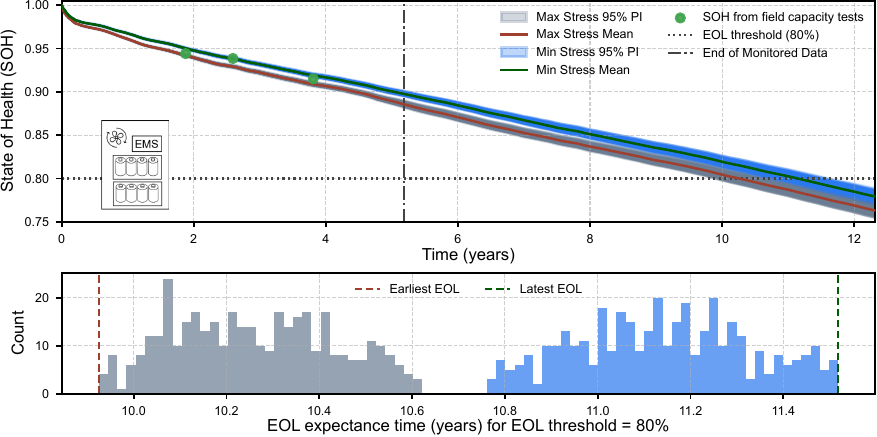}
  \caption{Prediction result for the sensitivity analysis. A synthetic calendar-aging model is retrained using an adapted dataset with reduced measurement spread at the dominant \(25^\circ\mathrm{C}\), 100\% SOC test point. Compared to the original model (Fig.~\ref{fig:Figgener14results}), the resulting prediction intervals are narrower, yielding a markedly more confident field prediction and confirming that the increased capacity loss of a single cell at this test point (Fig.~\ref{fig:outlier_calendar_point}) is a major driver of the aleatoric component of the field prediction uncertainty. This retrained model is used for this diagnostic experiment only and does not replace the framework's reported field predictions in section~\ref{sec:cs_results}. The epistemic uncertainty from limited laboratory coverage (section~\ref{sec:epistemic_mismatch}) remains unaffected.}  
  \label{fig:Figgener14results_reduced}
\end{figure}

The pronounced measurement spread observed at the \(25^\circ\mathrm{C}\), 100\% SOC calendar test point (section~\ref{sec:aleatoric_spread}) is a plausible dominant driver of aleatoric uncertainty in the field prediction, since the field operating profile spends substantial time in exactly this stress region. To test this hypothesis in isolation, we conduct a targeted sensitivity experiment. The capacity measurements of the single cell responsible for the unusually wide spread at this test point are replaced with a synthetic set implying a reduced spread of approximately 0.8\% (see the "Synthetic Cell 1 Data" in Fig.~\ref{fig:outlier_calendar_point}). All other test points and modeling choices are left unchanged. A calendar-aging model is retrained on this modified dataset for diagnostic purposes only and does not replace the model underlying the results reported in section~\ref{sec:cs_results}. Fig.~\ref{fig:Figgener14results_reduced} shows the effect of applying this retrained model to the same field operating profile. The resulting EOL uncertainty range narrows from 3 years (original model, section 3.2.3) to a little more than 1.5 years, a reduction of roughly 50\%. The earliest expected threshold-crossing shifts from 9 to almost 10 years, and the latest from 12 to 11.5 years, yielding a markedly more confident field prediction. This confirms that the high measurement spread at this single, frequently-visited test point is a major contributor to the aleatoric component of the field prediction uncertainty. With three measured cells per test point, this study cannot statistically distinguish whether "Cell 1" is a true outlier or whether the observed spread reflects genuinely large aleatoric variability at this test condition. Regardless of which explanation holds, the ambiguity itself illustrates how the design of laboratory aging studies constrains the confidence of downstream predictions: a larger database of replicate cells could reliably characterize whether the observed spread reflects the true variability of the aging process. Whether to treat "Cell 1" as an outlier is therefore not a purely technical choice but a risk decision: excluding it narrows EOL predictions, at the risk of understating true variability if the spread is instead genuine. The framework makes this trade-off explicit, leaving the choice and its risk to the party setting warranty or maintenance terms. This localized ambiguity aside, the epistemic uncertainty from limited laboratory coverage discussed in section~\ref{sec:epistemic_mismatch} would remain until that coverage is extended (see section~\ref{sec:implicaitons}).

\subsection{Implications for Laboratory Test Design}
\label{sec:implicaitons}
These insights give laboratory test designers actionable guidance. Because laboratory aging campaigns are time- and resource-intensive, their design must balance accelerated degradation observation against representative coverage of real operating conditions. Table~\ref{tab:testpoints} summarizes field-informed additions to the laboratory test matrix that could improve support for HESS operation characteristics and thereby reduce epistemic uncertainty in future applications. Accelerated high-stress test points remain essential to generate sufficient degradation within feasible experimental time. However, they should be complemented by field-relevant operating conditions that improve support in the regions most frequently occupied in real-world operation. Run in parallel with the existing test matrix for the same campaign duration, these conditions add no additional calendar time to the study, even though they will show correspondingly slower capacity loss. Even a small observed degradation signal anchors the local trajectory shape at these conditions in real data rather than in cross-condition generalization, providing a more reliable basis for temporal extrapolation beyond the observed measurement window. The primary cost of these additions is therefore the number of parallel test channels required, not experimental duration. In this way, future laboratory studies can preserve degradation observability while improving the reliability and calibration of probabilistic SOH and EOL predictions. Given the resource cost of testing additional cells, a full increase in replication across the entire matrix is unlikely to be practical. Instead, we recommend prioritizing higher cell counts at the field-dominant test points, like the ones identified in section~\ref{sec:aleatoric_spread} for this HESS use case. This would offer a broader statistical basis at the points that matter most. Specifically, doubling replication from three to six cells would markedly improve confidence in the measured aleatoric spread itself. The relative uncertainty of this spread estimate scales as $1/\sqrt{2(n-1)}$ and would fall from approximately 50\% at n=3 to 32\% at n=6. This would improve the ability to distinguish a true outlier from genuine variability. This targeted approach improves statistical confidence for the operating regimes represented in this manufacturer's specific usage scenario. Points outside these regions retain the original replication level and would require separate justification should future applications shift the dominant operating conditions.

\begin{table}[h]
    \centering
    \caption{Priority additions to the laboratory aging test matrix for HESS applications}
    \label{tab:testpoints}
    \begin{tabular}{p{0.33\linewidth} p{0.56\linewidth}}
        \toprule
        \textbf{Stress domain} & \textbf{Recommended addition} \\
        \midrule
        Low-rate cyclic operation &
        More granular charge and discharge tests below $0.4$C \\
        
        Temperature support &
        Denser test points in the field-relevant temperature range between $15\,^\circ\mathrm{C}$ and $35\,^\circ\mathrm{C}$  with optional extension to low-temperature outdoor scenarios \\
        
        SOC / DOC support &
        Shallow cycling tests in representative high- and low-SOC regions \\
        
        Dynamic operation &
        Dynamic load profiles with intermittent rest phases instead of only static stress points \\
        \bottomrule
    \end{tabular}
\end{table}

\section{Conclusion and Outlook}
\label{outlook}
This study demonstrates that a laboratory-trained, cell-level probabilistic aging model can be transferred to system-level SOH and EOL prediction for field-operated BESS. The 0.4\% mean absolute SOH error achieved on independent cell-level validation profiles demonstrates the framework's high accuracy. The cell-to-system approximation, which combines the Minimum- and Maximum-Stress Scenarios derived from field operating data, brackets the observed system-level degradation well: all three available field capacity measurements over approximately four years of operation fall within the predicted intervals. This indicates that the cell-to-system approximation provides a valid envelope for real-world degradation uncertainty. Notably, the two bounding scenarios differ only moderately in operational stress (temperature spread of roughly 5\%, current spread of roughly 9\%), yet their mean predicted end-of-life differs by 10 months. Once trajectory-level uncertainty in both scenarios is accounted for, the combined predicted end-of-life range spans approximately three years, about a third of the predicted system lifetime. This illustrates how sensitive long-term lifetime expectations are to variation in system-level operating conditions.  

Beyond this, the uncertainty-driver analysis provides a second, independently useful result: it explains why prediction intervals for the field operation profiles are wider than those obtained for the validation profiles. The alignment metrics comparing laboratory and field stress-factor distributions (section~\ref{sec:epistemic_mismatch}) reveal systematic coverage gaps, consistent with an increased epistemic contribution. The empirical spread of capacity measurements at dominant laboratory test points (section~\ref{sec:aleatoric_spread}) points to an increased aleatoric contribution. The sensitivity experiment makes this second effect concrete: replacing the elevated-degradation measurements of a single cell at one frequently visited test condition narrows the predicted EOL range by roughly 50\%.

Taken together, these contributions give the framework value beyond a single case study. Prior deterministic degradation estimators offer BESS manufacturers a single point estimate for warranty design, and even where probabilistic models exist, they are rarely validated against field capacity measurements (Table~\ref{tab:lit_review_calendar}). Here, EOL probability distributions are instead benchmarked directly against field capacity measurements, giving manufacturers a defensible basis for setting SOH-based warranty terms and understanding the associated risk exposure. This aligns with the growing regulatory emphasis on transparent SOH reporting. For operators and asset managers, estimating a deployed system's remaining lifetime has so far required either a costly, downtime-inducing on-site capacity test or an unverified assumption from datasheet specifications. The EOL probability distributions derived from logged data instead support risk-aware maintenance scheduling, replacement planning, and second-life decisions that affect a system's long-term reliability and economic return. For laboratories designing future aging campaigns, test-matrix coverage has traditionally relied on experience rather than a quantitative link to downstream prediction confidence. The alignment metrics and sensitivity analysis introduced here instead provide a transferable basis for prioritizing test resources. Finally, because the framework, code, and training data are openly available, other groups can apply the methodology to different chemistries, topologies, and field datasets.

Some limitations remain. As noted in section~\ref{sec:pred_framework}, the training data covers degradation observations only down to approximately 80\% SOH, so late-life knee-point behavior remains outside the scope of the present models. The cell-to-system approximation captures only extreme stress-driven bounds on degradation but does not yet model systematic effects such as current spreads within the pack or manufacturing-related cell-to-cell variability in initial capacity. 

Future work could partly close systematic coverage gaps caused by accelerated laboratory test conditions. One approach could be the generation of synthetic unaccelerated test points, for example using Arrhenius-based temperature scaling, to extend training data into field-relevant but slowly degrading regimes without proportional increases in test duration~\cite{Paarmann.2024}. Physics-informed neural networks that embed known degradation mechanisms directly into the model structure offer a complementary route to reducing reliance on purely data-driven extrapolation into sparsely sampled regions. Automated outlier-detection methods for training-data curation would help identify atypical cells such as the one flagged in the sensitivity analysis before they inflate predicted uncertainty. Finally, extending the benchmarking to additional field case studies and system topologies will establish the generality of the proposed framework.

By transferring cell-level laboratory characterization to system-level degradation prediction for field-operated systems, this framework moves uncertainty-aware battery lifetime prediction from a methodological demonstration toward a practically usable tool for warranty design, maintenance planning, and future test-campaign design.
% Use the style of numbering in square brackets.
% If nothing is used, default style will be taken.
%\begin{enumerate}[a)]
%\item 
%\item 
%\item 
%\end{enumerate}  

%\begin{itemize}
%\item 
%\item 
%\item 
%\end{itemize}  

%\begin{description}
%\item[]
%\item[] 
%\item[] 
%\end{description}  

%\clearpage %%Remove this from your manuscript

%

% Uncomment and use as the case may be
%\begin{theorem} 
%\end{theorem}

% Uncomment and use as the case may be
%\begin{lemma} 
%\end{lemma}
\section*{Funding and acknowledgment}
This research is funded by the German Federal Ministry for Economic Affairs and Energy (BMWE) via the research project BattLifeBoost (grant number 03EI4068C). The project is overseen by Project Management Juelich (PtJ).

\printcredits

\section*{Data availability}
The code and data supporting the findings of this study are openly available in the "BattProDeep" repository at \url{https://github.com/hsk-ses/BattProDeep}~\cite{BattProDeep.github}. The field dataset used for benchmarking was published by Figgener et al.~\cite{Figgener.2024} and is publicly available. The laboratory cell aging datasets used for model training were published by Naumann et al.~\cite{Naumann.2018, Naumann.2020} and are publicly available.

\appendix
\section{Cumulative Loss and Virtual Time Principle}
\label{appendix:cum_loss_virt_time}
\begin{figure}[tb]
    \centering
    \includegraphics[width=1\columnwidth]{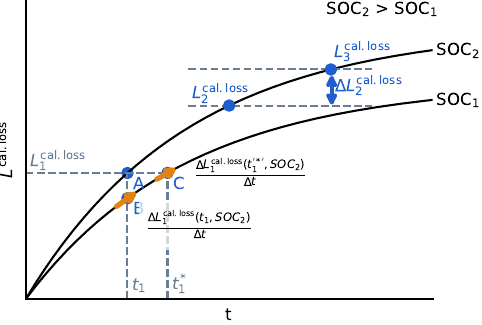} 
    \caption{Visualization of the principles of incremental loss and virtual time applied to distinguish between different stress states in dynamic usage profiles (indicated by different SOC levels here).  In the example shown, a transition happens between storage at $\mathrm{SOC_2}$ and storage at $\mathrm{SOC_1}$ at timestep $t_1$. To identify the correct loss rate at point C with a new stress level, the total loss at point A needs to be considered, encompassed in the virtual time $t^*$. The incremental loss is derived from the total losses of two successive timesteps (here $L_2^{\mathrm{cal. loss}}$ and $L_3^{\mathrm{cal. loss}}$) with identical stress states. All incremental losses contribute to the total degradation trajectory of dynamic load profiles. }
    \label{fig:inc_loss_vt}
\end{figure}

Fig.~\ref{fig:inc_loss_vt} depicts how for calendar aging the timeline is split into increments of $\Delta t$ and incremental losses $\Delta L$ are identified for the varying stress states (varying SOC levels in Fig.~\ref{fig:inc_loss_vt}) that dominate this time increment. Each portion of relative incremental loss is appended to the final dynamic loss trajectory. 

The virtual time $t^*$ represents the time duration that would have needed to pass to reach the level of calendar capacity loss reached in point A under the stress factors present in point B.
\section{HESS Operational Data}
\label{appendix:operational_data}
\begin{figure}[htb]
    \centering
    \includegraphics[width=1\columnwidth]{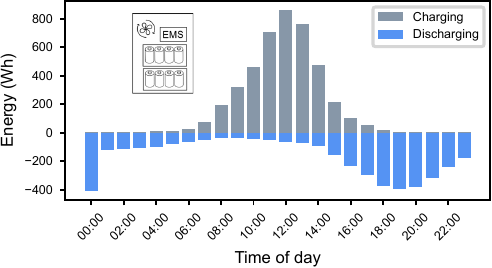} 
    \caption{Average hourly energy throughput (Wh) across the LFP-based field systems, indicating the magnitude of energy charged into and discharged from the HESS throughout a typical day.}
    \label{fig:energy_throughput}
\end{figure}
Fig.~\ref{fig:energy_throughput} illustrates the average hourly energy throughput across the LFP-based HESS systems published by Figgener et al.~\cite{Figgener.2024}. 
\begin{figure}[htb]
    \centering
    \includegraphics[width=1\columnwidth]{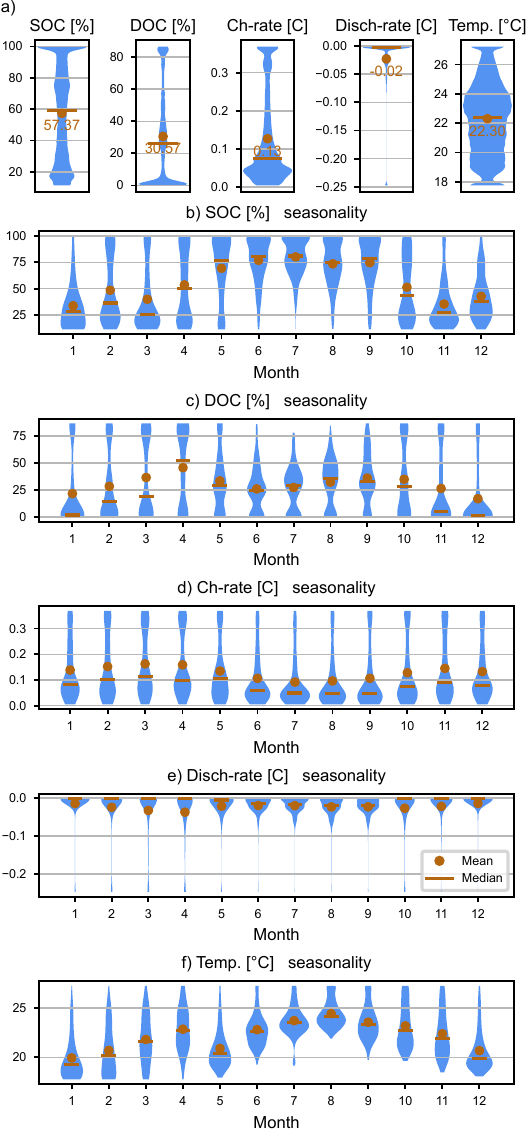} 
    \caption{Battery stress factor distributions derived from field operation of the relevant LFP systems in~\cite{Figgener.2024}. Plot a) shows the general distributions whereas plots b)-f) show distributions clustered by months to outline seasonal characteristics in the HESS operation. }
    \label{fig:field_violins_seasonal}
\end{figure}
All stress factors exhibit pronounced seasonality (see Fig.~\ref{fig:field_violins_seasonal} subplots b)-f)). This leads to more frequent full charging events and extended periods at high SOC during summer, as systems spend many hours at full charge and might not get fully discharged frequently. Conversely, winter conditions with lower PV generation result in fewer full cycles. During spring and fall, the SOC distribution is more balanced as the systems reach both full and empty states regularly, indicating regular full cycles. Similar seasonal effects can be observed for the DOC. The charge and discharge rate distributions similarly show slightly increased means during spring and fall and are lowest in winter with low PV generation and less charging. The battery housing temperatures follow seasonal ambient temperature patterns in Germany. Summer operation shows increased mean temperatures due to the combination of warm room temperatures and high charging power.

\section*{Declaration of competing interest}
The authors declare that they have no known competing financial interests or personal relationships that could have appeared to influence the work reported in this paper.

\section*{Declaration of generative AI use}
During the preparation of this work, the authors used Claude Code to improve the language and readability of the document. After using this tool/service, the authors reviewed and edited the content as needed and take full responsibility for the content of the published article.

\bibliographystyle{model1-num-names}
%\bibliographystyle{cas-model2-names}

\bibliography{BLB}



\end{document}